\documentclass[a4paper,fleqn]{cas-dc}
\pdfoutput=1 

\usepackage{ucs}
\usepackage{amssymb}
\usepackage{amsmath}
\usepackage{bbm}
\usepackage[greek,english]{babel}
\makeatletter
\newcommand*{\rom}[1]{\expandafter\@slowromancap\romannumeral #1@}
\makeatother
\usepackage{multirow}
\usepackage{longtable}
\usepackage{ragged2e}
\usepackage{graphicx}
\usepackage{dcolumn}
\usepackage{cuted}
\usepackage{bm}
\usepackage{hyperref}
\usepackage[square, comma, sort&compress, numbers]{natbib}
\usepackage[outdir=./]{epstopdf}

\def\tsc#1{\csdef{#1}{\textsc{\lowercase{#1}}\xspace}}
\tsc{WGM}
\tsc{QE}

\newcommand{\sixj}[6]
{
  \left\{ \begin{array}{ccc} #1 & #2 & #3 \\
    #4 & #5 & #6 \end{array} \right\}
}

\begin{document}
\let\WriteBookmarks\relax
\def\floatpagepagefraction{1}
\def\textpagefraction{.001}

\shorttitle{Application of Judd-Ofelt theory extension}    

\shortauthors{Hovhannesyan\textit{ et al.}}  

\title [mode = title]{Extension of Judd-Ofelt theory: Application on Eu\texorpdfstring{$^{3+}$}{Lg}, Nd\texorpdfstring{$^{3+}$}{Lg} and Er\texorpdfstring{$^{3+}$} {Lg}}  



%
\author[]{Gohar Hovhanensyan}[type=editor,
       orcid=0000-0003-4686-398X,
]




\ead{gohar.hovhannesyan@u-bourgogne.fr}



\affiliation{organization={Laboratoire Interdisciplinaire Carnot de Bourgogne, UMR 6303 CNRS-Univ.~Bourgogne Franche-Comt\'{e}},
            addressline={9 Avenue Alain Savary, BP 47 870}, 
            city={Dijon cedex},
            postcode={F-21078 },
            country={France}}

\author{Vincent Boudon}

\author{Maxence Lepers}









\begin{abstract}
We present a modified version of the Judd-Ofelt theory, which describes the intensities of f-f transitions for trivalent lanthanide ions (Ln$^{3+}$) in solids. In our model, the properties of the dopant are calculated with well-established atomic-structure techniques, while the influence of the crystal-field potential is described as a perturbation, by three adjustable parameters. Compared to our previous work [G. Hovhannesyan \textit{et al.}, J. Lumin. \textbf{241}, 118456 (2022)], the spin-orbit interaction within the first excited configuration 4f$^{w-1}$5d is described in a perturbative way, whereas it is exactly taken into account in the ground configuration 4f$^w$, using all the eigenvector components of the free-ion levels.
Moreover, the wavelength-dependence of the refractive index of the host material is also accounted for. We test the validity of our model on three ions: Eu$^{3+}$, Nd$^{3+}$ and Er$^{3+}$. The results of the extension are satisfactory, we are able to give a physical insight into all the transitions within the ground electronic configuration, and also to reproduce quantitatively experimental absorption oscillator strengths. We also performed calculations of standard JO parameters, and the results are in good agreement with the values reported in the literature. The code used to make the calculations is available on GitLab.
\end{abstract}

\maketitle

\section{Introduction}
\label{sec:introduction}

The Judd-Ofelt (JO) theory has been successfully applied since almost 60 years, to interpret the intensities of absorption and emissions lines of crystals and glasses doped with trivalent lanthanide ions (Ln$^{3+}$) \cite{judd1962, ofelt1962, hehlen2013}. Despite its remarkable efficiency for many cases, the standard version of the JO theory cannot reproduce some of the observed transitions, because of its strong selection rules. In order to overcome this issue many people tried to introduce extensions of the theory. This includes \textit{e.g.}~J-mixing \cite{tanaka1994, kushida2002, kushida2003}, the Wybourne-Downer mechanism \cite{downer1988, burdick1989}, velocity-gauge expression of the electric-dipole (ED) operator \cite{smentek1997}, relativistic or configuration-interaction (CI) effects \cite{smentek2000, smentek2001, wybourne2002, ogasawara2005, dunina2008}, purely \textit{ab initio} intensity calculations \cite{wen2014}. But despite all these improvements, even the most recent experimental studies use the standard version of JO theory \cite{ciric2019a, ciric2019b}.

In the standard version of the theory, a given transition can be characterized by line strengths, which are linear combinations of three parameters $\Omega_\lambda$ ($\lambda$ = 2,4,6), called JO parameters and adjusted by least-square fitting. Their formal expression depend on the crystal parameters as well as the properties of the Ln$^{3+}$ ion. But once $\Omega_\lambda$ values are obtained from a fit, it is not possible to separate the contributions of the crystal and of the ion. However much progress was done in recent years on the spectroscopy of free Ln$^{3+}$ ions \cite{wyart2007, radziute2015, meftah2016, meftah2017, freidzon2018, arab2019, gaigalas2019, chikh2021}, which makes it possible to use their properties as fixed parameters of a model similar to the JO one.

In a previous article \cite{hovhannesyan2022transition} (henceforth called Paper \rom{1}), we presented an extension of the JO theory, in which the free-ion properties are computed using Cowan's suite of codes \cite{rcecowan, mcguinness_cowan}, which allowed us to relax some of the strong assumptions. The calculated line strengths are linear combinations of three adjustable parameters, which are only functions of the crystal-field potential. In Paper \rom{1}, the spin-orbit interaction of the ion is treated using perturbation theory, both in the ground and in the first excited configurations.

In the present article by contrast, the spin-orbit interaction is fully taken into account in the ground configuration. We include all the eigenvector components of a given level, while in the previous version only the four leading ones were included. We also account for the wavelength dependence of the host material refractive index, using the modified Sellmeier equation. We test the validity of our new model on three ions: Eu$^{3+}$, Nd$^{3+}$ and Er$^{3+}$. The performance are similar to the standard JO model, but in addition, we are able to interpret some transitions which are strictly forbidden in the JO theory.

The article is organized as follows. In section \ref{sec:model} we describe our new extension of the JO theory, in particular how the line strength is modified with respect to Paper \rom{1} (see subsection \ref{subsec:extension_description}). We perform free-ion calculations and apply the theory on Eu$^{3+}$, Nd$^{3+}$ and Er$^{3+}$ (see subsections \ref{sec:Eu3+}, \ref{sec:Nd3+} and \ref{sec:Er3+}, respectively). And, finally, section \ref{sec:conclusions} contains conclusions and prospects for the work.

\section{Description of the model}
\label{sec:model}

The aim of the JO theory and of its extension is to calculate line intensities of transitions between levels belonging to the lowest electronic configuration 4f$^w$ of lanthanide ions Ln$^{3+}$ placed in a crystal or solid environment. The calculated intensities are adjusted using least-square fitting with experimental values, most often of the absorption oscillator strengths. Using the fitted parameters, other quantities like the Einstein coefficient for spontaneous emission can also be predicted. Oscillator strengths and Einstein coefficients are proportional to the transition line strength, whose calculation is described in subsection \ref{subsec:extension_description}. These calculated values are used in a least-square fitting procedure, see subsection \ref{subsec:lsf}, in which the experimental line strengths are calculated from measured oscillator strengths with wavelength-dependent refractive indices, as described in subsection \ref{subsec:refr_index}.

\subsection{Calculation of line strengths}
\label{subsec:extension_description}

In the standard JO theory, the electric-dipole (ED) line strength $\mathcal{S}_\mathrm{ED}$ is a linear combination of three adjustable quantities $\Omega_\lambda$ with $\lambda = 2$, 4 and 6, which are functions of free-ion properties like energies and transitions integrals, and of the crystal-field (CF) parameters $A_{kq}$ characterizing the potential energy created by the host material as follows
\begin{equation}
  V_\mathrm{CF} = \sum_{kq} A_{kq} P_q^{(k)}
  \label{eq:vcf}
\end{equation}
where $P_q^{(k)}$ is the electric-multipole tensor operator of rank $k$ and component $q$. The formal expression of $\Omega_\lambda$ JO parameters is established using time-independent quantum perturbation theory up to second order \cite{walsh2006}, assuming that the CF potential induces a weak coupling between the lowest configuration 4f$^w$  and the first excited one 4f$^{w-1}$5d, responsible for the activation of the ED transitions.

In paper I, we propose an extension of the standard JO theory in which the free-ion properties are not treated as adjustable parameters, but calculated using well-established techniques of atomic-structure calculations. The line strength is also a linear combination of three adjustable parameters $X_k$ ($k=1$, 3 and 5), that only depend on the CF parameters,
\begin{equation}
  X_k = \frac{1}{2k+1} \sum_{q=-k}^k \left|A_{kq}\right|^2 .
  \label{eq:xk}
\end{equation}
Unlike the standard and most common extensions of the JO model, we do not introduce effective operators, like the so-called unit-tensor operator $U^{(\lambda)}$ \cite{cowan1981}, but rather work on the matrix elements of the CF and ED operators.

More specifically in paper I, we present two different calculations: (i) where the spin-orbit (SO) interaction within the 4f$^{w-1}$5d configuration is not included, and (ii) where it is included.
In version (i), the ED transition amplitude $D_{12}$ is calculated with the second-order perturbation theory in which the perturbation operator is $V_\mathrm{CF}$. The unperturbed states are the free-ion levels of the lowest configuration 4f$^w$. Therefore the 4f$^w$ SO interaction is fully accounted for, as it is part of the unperturbed Hamiltonian. In version (ii), the perturbation operator is $V_\mathrm{CF} + H_\mathrm{SO}$, and in order to catch the effect of both terms, $D_{12}$ is calculated with the third-order perturbation theory. Because $H_\mathrm{SO}$ is accounted for in a perturbative way both in the ground and the excited configurations, the unperturbed states are the free-ion manifolds, \textit{i.e.}~levels without SO interaction. In other words, all the $J$ levels inside a given manifold, like $^7$F$_J$ in Eu$^{3+}$, are degenerate.

In the present work, we merge the two previous versions as follows. We consider as unperturbed states the free-ion levels of the ground configuration written in pair coupling,
\begin{equation}
  |\Psi_i^0\rangle = \sum_{\alpha_i L_i S_i} c_{\alpha_i L_i S_i} 
    \left|n\ell^w\,\alpha_i L_i S_i J_i M_i \right\rangle \,,
  \label{eq:0th-lev-gr}
\end{equation}
where $i=1,2$ describes the lower and upper levels, and $L_i$, $S_i$, $J_i$, $M_i$ respectively denote the orbital, spin, total angular momenta and its $z$-projections. The indices $\alpha_i$, standing for the seniority numbers, are sometimes necessary to distinguish manifolds with the same $L_i$ and $S_i$ (for example $^5$D1, $^5$D2 and $^5$D3 in Eu$^{3+}$). The $c_{\alpha_i L_i S_i}$ coefficients are the eigenvector components of the ionic Hamiltonian in LS coupling scheme. Because for Ln$^{3+}$ ion in the lowest configuration, there is most often one dominant LS component (with $|c_{\alpha_i L_i S_i}|^2 > 0.7$), the free-ion levels are labeled with that component. In the present work, we take all the components into account, whereas in paper I we only took the four leading ones (due to practical reasons).

The transition amplitude $D_{12}$ is now the sum of the second-order contribution describing the bare influence of the CF, and a third-order contribution describing the influence of the CF and excited-configuration SO interaction (the so-called Downer-Wybourne mechanism). The expression of $D_{12}$ becomes
\begin{align}
  D_{12} = & \sum_{t} \left[ \frac{ \langle \Psi_1^0 | 
    V_{\mathrm{CF}} | \Psi_t^0 \rangle 
    \langle \Psi_t^0 | P_p^{(1)} | \Psi_2^0 \rangle} {E_1-E_t}
      \right. \nonumber \\
   & + \frac{ \langle \Psi_1^0 | P_p^{(1)}
     | \Psi_t^0 \rangle \langle \Psi_t^0 
     |V_{\mathrm{CF}} | \Psi_2^0 \rangle } 
     {E_2-E_t} \nonumber \\
   & + \sum_{u} \left\{ \frac{ \langle \Psi_1^0 | 
    V_{\mathrm{CF}} | \Psi_t^0 \rangle
    \langle \Psi_t^0 | H_\mathrm{SO} | \Psi_u^0 \rangle
    \langle \Psi_u^0 | P_p^{(1)} | \Psi_2^0 \rangle }
     {(E_1-E_t)^2} \right. \nonumber \\
   & + \left. \left. \frac{ \langle \Psi_1^0 | 
    P_p^{(1)} | \Psi_t^0 \rangle
    \langle \Psi_t^0 | H_\mathrm{SO} | \Psi_u^0 \rangle
    \langle \Psi_u^0 | V_{\mathrm{CF}} | \Psi_2^0 \rangle }
     {(E_2-E_u)^2} \right\} \right] \,,
  \label{eq:d12-1}
\end{align}
where $|\Psi_{t,u}^0\rangle = |n\ell^{w-1} \overline{\alpha} \overline{L} \overline{S}, \, n'\ell' L'_{1,2} S'_{1,2} J'M' \rangle$ are unperturbed LS states of the excited configuration: namely $n\ell = 4$f and $n'\ell' = 5$d. Note that the matrix elements of $V_\mathrm{CF}$ are functions of the one-electron radial integrals $\langle n' \ell' | r^k | n \ell \rangle = \int^{+\infty}_0 dr P_{n'\ell'}(r) r^k P_{n\ell}(r)$, with $k = 1$, 3 and 5, and ($P_{n\ell}$, $P_{n'\ell'}$) the wave function of the corresponding orbital. The component $p=0$ ($\pm 1$) of the dipole operator $P_p^{(1)}$ corresponds to $\pi$ ($\sigma^{\pm}$) light polarizations.

\newpage

\begin{strip}
\hspace{12pt} Employing the same angular-momentum properties as in paper I, we obtain for the transition amplitude
\begin{align}
  D_{12} & = \sum_{\alpha_{1}L_{1}S_{1}}c_{\alpha_{1}L_{1}S_{1}}
    \sum_{\alpha_{2}L_{2}S_{2}}c_{\alpha_{2}L_{2}S_{2}}
    \sum_{kq}\,A_{kq} \sum_{\lambda\mu} (-1)^{J_{1}+J_{2}-\lambda} 
    \sqrt{\frac{2\lambda+1}{2J_{1}+1}} C_{kq1p}^{\lambda\mu} 
    C_{J_{2}M_{2}\lambda\mu}^{J_{1}M_{1}} \nonumber \\
  & \times \sum_{J'} \left[ \sixj{k}{1}{\lambda}{J_2}{J_1}{J'}
   \left( \mathcal{D}_{12,J'}^{(k1)} + \mathcal{D}_{12,J'}^{(k01)} \right)
   + \left(-1\right)^{1+k-\lambda}
   \sixj{1}{k}{\lambda}{J_2}{J_1}{J'}
   \left( \mathcal{D}_{12,J'}^{(1k)} + \mathcal{D}_{12,J'}^{(10k)} \right) \right],
  \label{eq:d12-2}
\end{align}
where $C_{a\alpha b\beta}^{c\gamma}$ is a Clebsch-Gordan coefficient and the quantity between curly brackets is a Wigner 6-j symbol.
For the line strength $\mathcal{S}_\mathrm{ED} = \sum_{M_1 M_2 p} (D_{12})^2$, one has
\begin{align}
  \mathcal{S}_{\mathrm{ED}} = & \sum_{\alpha_{1a}L_{1a}S_{1a}} 
   c_{\alpha_{1a}L_{1a}S_{1a}} \sum_{\alpha_{2a}L_{2a}S_{2a}} c_{\alpha_{2a}L_{2a}S_{2a}} \sum_{\alpha_{1b}L_{1b}S_{1b}} c_{\alpha_{1b}L_{1b}S_{1b}} \sum_{\alpha_{2b}L_{2b}S_{2b}} c_{\alpha_{2b}L_{2b}S_{2b}} \sum_{kq}\frac{|A_{kq}|^{2}}{2k+1} 
  \nonumber \\
  \times & \sum_{J'} \left[ \frac{1}{2J'+1} \left( 
    \widetilde{\mathcal{D}}_{1a,2a,J'}^{(k1)}
      \widetilde{\mathcal{D}}_{1b,2b,J'}^{(k1)}
      + \widetilde{\mathcal{D}}_{1a,2a,J'}^{(1k)}
      \widetilde{\mathcal{D}}_{1b,2b,J'}^{(1k)}
    \right)
    + \sum_{J''} \left(-1\right)^{1+k+J'+J''} \right.
  \nonumber \\
 & \left. \phantom{\sum_{J''}}
   \times \left( \sixj{k}{J_1}{J'}{1}{J_2}{J''}
     \widetilde{\mathcal{D}}_{1a,2a,J'}^{(k1)}
     \widetilde{\mathcal{D}}_{1b,2b,J''}^{(1k)}
     + \sixj{1}{J_1}{J'}{k}{J_2}{J''}
     \widetilde{\mathcal{D}}_{1a,2a,J'}^{(1k)} 
     \widetilde{\mathcal{D}}_{1b,2b,J''}^{(k1)}
   \right) \right] \,.
  \label{eq:sed-1}
\end{align}
where $\widetilde{\mathcal{D}}_{12,J'}^{(k_1 k_2)} = \mathcal{D}_{12,J'}^{(k_1 k_2)} + \mathcal{D}_{12,J'}^{(k_1 0 k_2)}$, and $\mathcal{D}_{12,J'}^{(k_1 k_2)}$ and $\mathcal{D}_{12,J'}^{(k_1 0 k_2)}$ are given in Eqs.~(8), (9) and (23) of Paper I.
\end{strip}

Due to angular-momentum selection rules, these equations impose some conditions on the indices:
\begin{itemize}
  \item $|\ell-\ell'| \le k \le \ell+\ell'$ and $\ell + \ell' + k$ even, which gives $k=1$, 3 and 5, since $\ell = 3$ and $\ell' = 2$.
      \item $k-1 \le \lambda \le k+1$, which gives $\lambda = 0$ to 6. In the standard JO theory, one has $\lambda=k+1$.
  \item $|J_1-J_2| \le \lambda \le J_1+J_2$, which gives $0 \le |J_1-J_2| \le 6$.
  \item $0 \le |L_1-L_2| \le 7$.
  \item $|S_1-S_2| = 0$ or 1.
\end{itemize}

Regarding the last rule, the second-order correction, given by the two first lines of Eq.~\eqref{eq:d12-1}, imposes $|S_1-S_2| = 0$. Therefore spin change comes from the fact that the free-ion 4f$^w$ levels have different spin components $S_i$, even though one is by far dominant. The two last lines of Eq.~\eqref{eq:d12-1} may in contrast given $S_1-S_2 = \pm 1$ due to the SO interaction within the 4f$^{w-1}$5d configuration.

\subsection{Least-square fitting procedure}
\label{subsec:lsf}

Using the expression \eqref{eq:sed-1} for the ED line strength, we now seek to minimize the standard deviation between calculated and experimental line strengths
\begin{equation}
  \sigma = \left[ \frac{ \sum_{i=1}^{N_\mathrm{tr}} 
    \left( \mathcal{S}_{\mathrm{exp},i}
      - \mathcal{S}_{\mathrm{ED},i} \right)^2}
    {N_\mathrm{tr}-N_\mathrm{par} }  \right]^{\frac{1}{2}},
  \label{eq:sigma}
\end{equation}
where $N_\mathrm{tr}$ is the number of experimental transitions included in the calculation and $N_\mathrm{par}=3$ is the number of adjustable parameters.
The experimental line strengths in atomic units are given as function of the measured oscillator strengths $f_\mathrm{exp}$ by
\begin{equation}
  \mathcal{S}_\mathrm{exp} = \frac{3(2J_1+1) \hbar^2}{2 m_e a_0^2 (E_2 - E_1)}
    \frac{n_r}{\chi_\mathrm{ED}} f_\mathrm{exp}
  \label{eq:s-exp}
\end{equation}
where $n_r$ is the host refractive index and is dependent on wavelength and $\chi_\mathrm{ED} = (n_r^2+2)/9$ the local-field correction in the virtual-cavity model (see for example Ref.~\cite{aubret2018}).

It is convenient to give the so-called relative standard deviations, which is the ratio $\sigma/\mathcal{S}_\mathrm{max}$ between the standard deviation and the maximum value among the experimental oscillator strengths. It is often expressed as a percentage.

After the fitting, using these optimal $X_k$ parameters, we can predict line strengths, oscillator strengths and Einstein $A$ coefficients, for other transitions. Of course, that procedure only involves transitions with a predominant ED character; magnetic-dipole (MD) transitions like $^5$D$_0 \leftrightarrow {}^7$F$_1$ and $^5$D$_1 \leftrightarrow {}^7$F$_0$ are therefore excluded from the fit. For them, the MD line strength $\mathcal{S}_\mathrm{MD}$, oscillator strengths and Einstein coefficients can be calculated from the free-ion eigenvectors.

\subsection{Wavelength dependence of refractive index}
\label{subsec:refr_index}

The refractive index of a material depends on the optical frequency or wavelength; this dependence is called chromatic dispersion. Typical refractive index values for glasses and crystals in the visible spectral region are in the range from 1.4 to 2.8, and typically the refractive index increases for shorter wavelengths (normal dispersion). The wavelength-dependent refractive index of a transparent optical material can often be described analytically with Cauchy's equation, which contains several empirically obtained parameters. The most general form of Cauchy's equation is
\begin{equation}
   n_r(\lambda) = A + {\frac {B}{\lambda^{2}}} + {\frac{C}{\lambda^{4}}} + \cdots ,
  \label{eq:Cauchy}
\end{equation}
where $n$ is the refractive index, $\lambda$ is the wavelength, $A$, $B$, $C$, etc.~are coefficients that can be determined for a material by fitting the equation to measured refractive indices at known wavelengths.

The Sellmeier equation is a later development of Cauchy's work that handles anomalously dispersive regions, and more accurately models a material refractive index across the ultraviolet, visible, and infrared spectrum. In its original and the most general form, the Sellmeier equation is given by
\begin{equation}
  n_r^{2}(\lambda) = n_0^{2} + \sum_{i=1}^m {\frac {A_{i}\lambda^{2}}
    {\lambda^{2}-B_{i}}},  
  \label{eq:Sell_1}
\end{equation}
where $n_0$ is the refractive index in vacuum, $\lambda$ is the wavelength, and $A_i$ and $B_i$ are experimentally determined Sellmeier coefficients.
The literature contains a great variety of modified equations which are also often called Sellmeier formulas. A somehow general form, gathering the Sellmeier and Cauchy ones, and sometimes used in papers dealing with Ln$^{3+}$ ions, is as follows:
\begin{equation}
  n_r^{2}(\lambda) = n_0^{2}+\sum_{i=1}^m \frac{A_i\lambda^{2}}{\lambda^{2}-B_i}
    + \sum_{j=1}^p\frac{C_j}{\lambda^{2j}}
  \label{eq:Sell_2}
\end{equation}
However, in the experimental studies with which we deal here, the authors use the simple formula
\begin{equation}
  n_r^{2}(\lambda) = n_0^{2}+\frac{A\lambda^{2}}{\lambda^{2}-B}.
  \label{eq:Sell_3}
\end{equation}
obtained by setting $m=1$ and $p=0$.

\section{Results on europium}
\label{sec:Eu3+}

\begin{table*}
\caption{\label{tab:UK_Eu} Values of the reduced matrix elements of the squared unit-tensor operator $[U^{(\lambda)}]^2$ (from the present work) for the transitions of Eu$^{3+}$ present in Ref.~\cite{babu2000} (see subsection \ref{subsec:Eu-babu}), compared with the values reported in Ref.~\cite{carnall1968electronic} (Rep.).}
\begin{tabular}{rllllll}
\toprule
%
Transition & \multicolumn{2}{c}{$[U^{(2)}]^2$} & \multicolumn{2}{c}{$[U^{(4)}]^2$} & \multicolumn{2}{c}{$[U^{(6)}]^2$}  \\ \cline{1-7}
 & Our & Rep. & Our & Rep. & Our & Rep. \\
\cline{2-3} \cline{4-5} \cline{6-7} \\
$^7$F$_{1} \leftrightarrow {}^7$F$_{6}$ & 0 & 0 & 0 & 0 & 0.3772 & 0.3773 \\
$^7$F$_{0} \leftrightarrow {}^7$F$_{6}$ & 0 & 0 & 0 & 0 & 0.1449 & 0.1450  \\
$^7$F$_{1}\leftrightarrow {}^5$D$_{1}$ & 0.0026 & 0.0026 & 0 & 0 & 0 & 0 \\
$^7$F$_{0}\leftrightarrow {}^5$D$_{2}$ & 0.0008 & 0.0008 & 0 & 0 & 0 & 0 \\
$^7$F$_{1}\leftrightarrow {}^5$D$_{3}$ & 0.0004 & 0.0004 & 0.0013 & 0.0012 & 0 & 0 \\
$^7$F$_{1}\leftrightarrow {}^5$L$_{6}$ & 0 & 0 & 0 & 0 & 0.0096 & 0.0090 \\
$^7$F$_{0}\leftrightarrow {}^5$L$_{6}$ & 0 & 0 & 0 & 0 & 0.0147 & 0.0155 \\
$^7$F$_{0}\leftrightarrow {}^5$G$_{2}$ & 0.0006 & 0.0006 & 0 & 0 & 0 & 0 \\
$^7$F$_{0}\leftrightarrow {}^5$D$_{4}$ & 0 & 0 & 0.0013 & 0.0011 & 0 & 0\\
\bottomrule
\end{tabular}
\end{table*}

\subsection{Free-ion calculation}
\label{subsec:eu-free}

Our free-ion calculations are presented in Paper I for Eu$^{3+}$, Nd$^{3+}$ and Er$^{3+}$ and are recalled here. They require experimental energies for the ground and the first excited electronic configurations. For the Eu$^{3+}$ ground configuration 4f$^6$, we find them on the NIST ASD database \cite{NIST_ASD}. However, no experimental level has been reported for the 4f$^5 5d$ configuration. Because the 4f$^w$ configurations (with $2 \le w \le 12$) and the 4f$^{w-1}$5d ones (with $3 \le w \le 13$) possess the same energy parameters, we perform a least-square fitting calculation of some 4f$^{w-1}$5d configurations for which experimental levels are known, namely for Nd$^{3+}$ ($w=3$) and Er$^{3+}$ ($w=11$) \cite{wyart2007, meftah2016, arab2019, chikh2021}. Then, relying on the regularities of the scaling factors $f_X$ along the lanthanide series, we multiply the obtained scaling factors given in Table 1 of Paper I by the HFR parameters for Eu$^{3+}$ to compute the energies of 4f$^5$5d configuration.

The interpretation of Nd$^{3+}$ and Er$^{3+}$ spectra show that, because CI mixing is very low, a one-configuration approximation can safely be applied in both parities, which is done here. For Nd$^{3+}$, experimentally known levels are taken from the article of Wyart \textit{et al.}~\cite{wyart2007}. There are 41 levels for 4f$^3$ configuration and 111 for 4f$^2 5d$ configuration. For Er$^{3+}$, 38 experimental levels of the configuration 4f$^{11}$ and 58 of 4f$^{10} 5d$ are taken from Meftah \textit{et al.}~\cite{meftah2016}. For the 4f$^6$ configuration of Eu$^{3+}$, the NIST database gives 12 levels \cite{NIST_ASD}. Figure 1 of Paper I presents the calculated energy levels for the two lowest configurations.

In addition to the free-ion ED reduced matrix element ($k=1$), our model requires those for $k=3$ (octupole) and $k=5$, which depend on the radial transition integral $\langle \mathrm{4f} | r^k | \mathrm{5d} \rangle$. We have calculated those integrals with a home-made Octave code, reading the HFR radial wave functions $P_{4f}$ and $P_{5d}$ computed by Cowan's code RCN. We obtain 1.130629$\,a_0$, -3.221348$\,a_0^3$ and 21.727152$\,a_0^5$ for $k=1$, $k=3$ and $k=5$, respectively, while the $k=1$ value calculated by Cowan is 1.130618$\,a_0$.

We have also calculated the reduced matrix elements of the so-called doubly reduced unit tensor operators of rank $k$ of Eu$^{3+}$, $[U^{(\lambda)}]^2$, which appear in the standard JO theory and are independent of the crystal host. This allows us to test the quality of our free-ion calculation. In this respect, Table \ref{tab:UK_Eu} shows a very good agreement between our values and those from the seminal article of Carnall \cite{carnall1968electronic}. The transitions present in the table are those used in the fitting procedure with the data from Babu \textit{et al.}~\cite{babu2000} (see next subsection).

\subsection{Eu$^{3+}$ in lithium fluoroborate}
\label{subsec:Eu-babu}

Now we will benchmark our model with two sets of experimental data. The first one comes from the thorough investigation of Babu \textit{et al.} \cite{babu2000}. In that article they measure absorption oscillator strengths and interpret them with standard JO theory. Their study deals with transitions within the ground manifold $^7$F and between the ground and excited manifolds $^5$D, $^5$L and $^5$G for Eu$^{3+}$-doped lithium borate and lithium fluoroborate glasses. We focus on the oscillator strengths given in their Table 3. We have taken the Sellmeier coefficients $n_0=1$, $A=1.2428$, $B=0.023833~\mu$m$^2$, of the host refractive index from Adamiv \textit{et al.} \cite{adamiv2011optical}, where optical properties of borate glasses have been measured.

With the standard JO theory applied to 9 transitions, we find a relative standard deviation \eqref{eq:sigma} of 8.52~\%. With our model \eqref{eq:sed-1}, we find 8.19~\% by assuming a wavelength-independent refractive index, and 8.03~\% by applying the Sellmeier equation \ref{eq:Sell_3}. Therefore our model has slightly better performance, especially when we include the dispersion in the host material.

We have also investigated the effect of dispersion on the optimal JO parameters. When including the wavelength-dependence, all of them decrease: $\Omega_2$ from 25.79$\times 10^{-20}$cm$^2$ to 18.73$\times 10^{-20}$ cm$^2$, $\Omega_4$ from 17.88$\times 10^{-20}$ cm$^2$ to 12.58$\times 10^{-20}$ cm$^2$ and, finally, $\Omega_6$ from 3.015$\times 10^{-20}$ cm$^2$ to 2.253$\times 10^{-20}$ cm$^2$, making the comparison with values reported in Babu \textit{et al.} \cite{babu2000} better (see table \ref{tab:JO_Eu}).

This table also presents the optimal fitted parameters $X_k$ of our extension in atomic units, that is to say $(E_h/a_0^k)^2$ with $E_h$ the Hartree energy. It is difficult to compare them directly with the $\Omega_\lambda$ parameters because they do not represent the same quantity, but one can say they follow similar trends, namely $\Omega_2 > \Omega_4 > \Omega_6$ and $X_1 > X_3 > X_5$.

\begin{figure}
\includegraphics[scale=0.63]{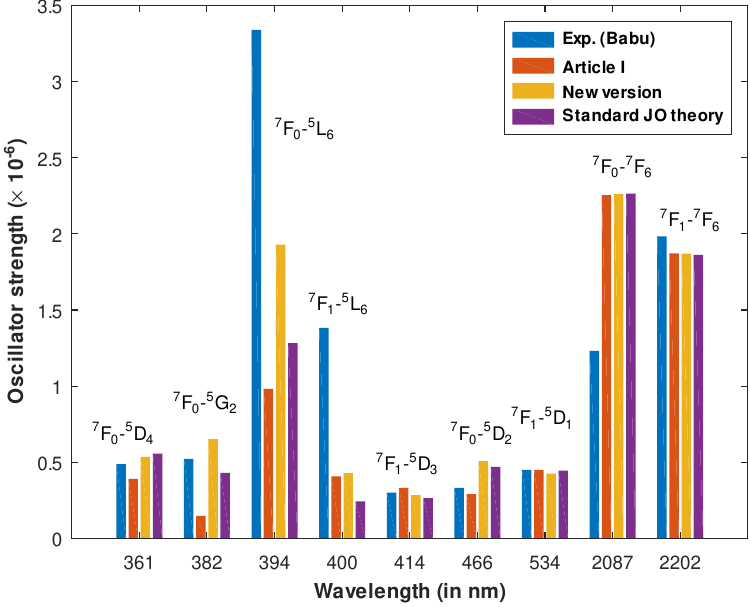}
\includegraphics[scale=0.63]{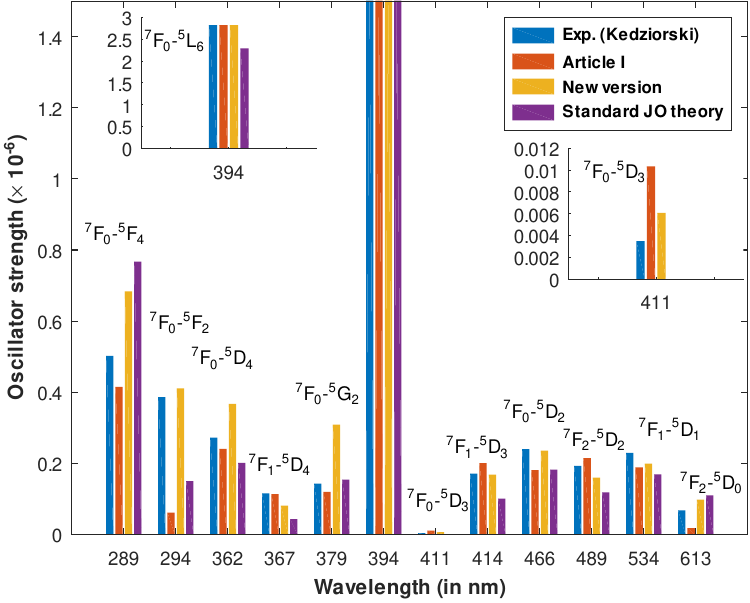}
\caption{\label{fig:Eu_jo_histogram} Comparison between experimental (top panel: Li fluoroborate \cite{babu2000}, bottom panel: in acetate \cite{kedziorski2007}) and theoretical (3rd order correction of article \rom{1} and new versions) oscillator strengths of absorption, plotted as function of the transition wavelength (not at scale) Eu$^{3+}$. The transitions are labeled with the LS-term quantum numbers of the Eu$^{3+}$ free ion.
}
\end{figure}

\begin{table*}
\centering
\caption{\label{tab:JO_Eu} Values of Judd-Ofelt parameters (in 10$^{-20}$ cm$^2$) and $|A_{kq}|^2$ (in a.u.) for Eu$^{3+}$ from the present work (Our), compared to values reported in the literature (Rep.). The experimental oscillator strengths and Judd-Ofelt parameters from Babu \textit{et al.} \cite{babu2000} are from set B (with thermal corrections). Judd-Ofelt parameters are calculated with the transition set from Kedziorski \textit{et al.} \cite{kedziorski2007}.}
\begin{tabular}{lccccccccc}
\toprule
%
 & $X_1$ & \multicolumn{2}{c}{$\Omega_2$} & $X_3$ & \multicolumn{2}{c}{$\Omega_4$} & $X_5$ & \multicolumn{2}{c}{$\Omega_6$} \\
 & ($10^{-4}$\,a.u.) & \multicolumn{2}{c}{($10^{-20}$\,cm$^2$)} & ($10^{-5}$\,a.u.) & \multicolumn{2}{c}{($10^{-20}$\,cm$^2$)} & ($10^{-8}$\,a.u.) & \multicolumn{2}{c}{($10^{-20}$\,cm$^2$)} \\
 \cline{2-10}
 & & Our & Rep. & & Our & Rep. & & Our & Rep. \\
\cline{3-4} \cline{6-7} \cline{9-10} \\
Eu$^{3+}$ in Li fluoroborate \cite{babu2000} & 1.816 & 18.73 & 17.96 & 1.898 & 12.58 & 11.92 & 6.882 & 2.253 & 2.13 \\
Eu$^{3+}$ in acetate \cite{kedziorski2007} & 0.7887 & 6.991 & -  & 0.1317 & 8.326 & - & 0.1008 & 4.940 & - \\

\bottomrule
\end{tabular}
\end{table*}

\begin{table}
\caption{\label{tab:Eu_judd_ofelt}{Transition labels and ratios between theoretical and experimental oscillator strength for the third-order correction of Paper \rom{1} and for the present model for Eu$^{3+}$, when the experimental data are taken from \cite{babu2000} (second and third columns), and \cite{kedziorski2007} (two last columns). The last line presents the relative standard deviations for each model.}}

\begin{tabular}{ccccc}
\toprule
 & \multicolumn{2}{c}{Eu$^{3+}$ in Li fluoroborate} & \multicolumn{2}{c}{Eu$^{3+}$ in acetate} \\
Transition & \multicolumn{2}{c}{Babu \cite{babu2000}} & \multicolumn{2}{c}{Kedziorski \cite{kedziorski2007}} \\
\cline{2-3}
\cline{4-5}
Label & Paper \rom{1} & Present & Paper \rom{1} & Present \\ \hline
 & & & & \\
 ${}^7$F$_0\leftrightarrow {}^7$F$_4$ & & & 0.82 & 1.36 \\
 ${}^7$F$_0\leftrightarrow {}^7$F$_2$ & & & 0.16 & 1.06 \\
 ${}^7$F$_0\leftrightarrow {}^5$D$_4$ & 0.80 & 1.10 & 0.88 & 1.35 \\
 ${}^7$F$_1\leftrightarrow {}^5$D$_4$ & & & 0.98 & 0.70 \\
 ${}^7$F$_0\leftrightarrow {}^5$G$_2$ & 0.28 & 1.25 & 0.83 & 2.16 \\ 
 ${}^7$F$_0\leftrightarrow {}^5$L$_6$ & 0.29 & 0.58 & 0.99 & 1.00 \\
 ${}^7$F$_1\leftrightarrow {}^5$L$_6$ & 0.29 & 0.31 & \\
 ${}^7$F$_0\leftrightarrow {}^5$D$_3$ & & & 2.96 & 1.74 \\
 ${}^7$F$_1\leftrightarrow {}^5$D$_3$ & 1.10 & 0.94 & 1.18 & 0.98 \\
 ${}^7$F$_0\leftrightarrow {}^5$D$_2$ & 0.88 & 1.53 & 0.75 & 0.98 \\
 ${}^7$F$_2\leftrightarrow {}^5$D$_2$ & & & 1.11 & 0.83 \\
 ${}^7$F$_1\leftrightarrow {}^5$D$_1$ & 1.00 & 0.95 & 0.82 &0.87 \\
 ${}^7$F$_0\leftrightarrow {}^5$D$_2$ & & & 0.26 & 1.44 \\
 ${}^7$F$_0\leftrightarrow {}^7$F$_6$ & 1.83 & 1.84 & \\
 ${}^7$F$_1\leftrightarrow {}^7$F$_6$ & 0.94 & 0.94 & \\
\bottomrule
$\sigma/\mathcal{S}_\mathrm{max}$ & 8.45~\% & 8.03~\% & 6.21~\% & 4.92~\%
\end{tabular}
\end{table}

At present, we investigate the agreement between theory and experiment for each transition included in the fit. From the data set of \cite{babu2000} we have excluded the three transitions that have a significant MD character, namely ${}^7$F$_1 \leftrightarrow {}^5$D$_0$, ${}^7$F$_0 \leftrightarrow {}^5$D$_1$ and ${}^7$F$_1 \leftrightarrow {}^5$D$_2$, but also ${}^7$F$_0 \leftrightarrow {}^5$D$_0$ which will be discussed in subsection \ref{subsec:00}, and ${}^7$F$_0 \leftrightarrow {}^5$G$_4$ for which we observed a large discrepancy in Paper I. For the 9 remaining transitions, the upper panel of figure \ref{fig:Eu_jo_histogram} presents, as functions of the wavelength but not at scale, histograms of the experimental and various calculated oscillator strengths, obtained with the standard JO model, our third-order correction of Paper I, and the current version.

Our present model show equal or better performance than the standard JO model, except for the $^7$F$_0 \leftrightarrow {}^5$G$_2$ transition. The same trend is observed between the present model and the one of Paper I, except for $^7$F$_0 \leftrightarrow {}^5$D$_2$ transition, see also Table \ref{tab:Eu_judd_ofelt}. Remarkably, the three models give significantly smaller oscillator strengths than the experimental ones, for the transitions involving the $^5$L$_6$ level. This could come from an inaccuracy in the free-ion eigenvector of this level, underlying the three models. However, such an overestimation of the OS is not visible on the bottom panel of figure \ref{fig:Eu_jo_histogram} with another data set. Another possible explanation is that those transitions overlap with ones involving another excited level close in energy. Note that, with the optimal $X_k$ parameters, we obtain an OS of $4.054 \times 10^{-7}$, corresponding a very satisfactory ratio of 1.10 with respect to the experimental value.

\subsection{Eu$^{3+}$ in acetate}

As a second data set we use absorption transitions from Kedziorski \textit{et al.} \cite{kedziorski2007}, where the authors present OSs for Eu$^{3+}$ in acetate crystal. We only consider resolved transitions between individual free-ion levels: namely we exclude those labeled $^7$F$_0\leftrightarrow {}^5$G$_{4,5,6}$ and $^7$F$_0\leftrightarrow {}^5$H$_{4,5,6}$. We also exclude the $^7$F$_1\leftrightarrow {}^5$D$_2$ due to its strong MD character, as well as the $^7$F$_0\leftrightarrow {}^5$D$_0$ one due to strong discrepancy.
To the best of our knowledge, there are no Sellmeier coefficients in the literature for acetate crystal, and the calculations were carried out under the assumption that the refractive index is constant and equal to 1.570.

Table \ref{tab:JO_Eu} presents the optimal fitting parameters $X_k$ and $\Omega_\lambda$. Comparison with literature values of standard JO parameters was not possible because in the article of Kedziorski \textit{et al.} \cite{kedziorski2007} these quantities are not discussed.

When 12 transitions are included in the fitting procedure, the relative standard deviation given by the JO model is 6.49~\%, the one given by the present model is 4.92~\%, while the standard deviation from article \rom{1} is 6.21~\%.
Transition $^7$F$_0\leftrightarrow {}^5$I$_4$ is excluded because our model overestimates the oscillator strength for this transition in comparison with the one mentioned in the article. Transition $^7$F$_0\leftrightarrow {}^5$I$_6$ is mentioned to have superimposed absorption bands with transition $^7$F$_0\leftrightarrow {}^5$H$_6$ in the article of Bukietynska \textit{et al.} \cite{bukietynska2001f}, on which most of the discussion in the article of Kedziorski \textit{et al.} is based. In order to avoid the possible confusion in identification of the peaks we exclude this transition from our fitting procedure.

A comparison between experimental and the OSs calculated with the standard JO model, the one resulting from article I and the one of the present article are shown in the bottom panel of figure \ref{fig:Er_jo_histogram}. The two insets are dedicated to the $^7$F$_0\leftrightarrow {}^5$L$_6$ and the $^7$F$_0\leftrightarrow {}^5$D$_3$ transitions which are not well visible on the main plot. In accordance with the relative standard deviations, our models systematically give better OSs than the standard JO one, except for the $^7$F$_0\leftrightarrow {}^5$G$_2$ transition. Note that the JO model cannot describe the $^7$F$_0\leftrightarrow {}^5$D$_3$ transition \cite{walsh2006, binnemans2015}, and that the present model gives a closer OS than the model of Paper I. For some transitions our present extension works better, while for others, the one of article I has better results, as shows Table \ref{tab:Eu_judd_ofelt}.

\subsection{The ${}^5$D$_0$-${}^7$F$_0$ transition}
\label{subsec:00}

The occurrence of the ${}^5$D$_0 \leftrightarrow {}^7$F$_0$ transition in Eu$^{3+}$ is a well-known example of the breakdown of the standard JO theory, due to its strong selection rule \cite{binnemans2015}. The most frequent explanations is the so-called $J$-mixing or the mixing of low-lying charge-transfer states into the eigenvector of the 4f$^6\,^7$F$_0$ ground level. $J$-mixing is due to the admixture of the $^7$F$_{2,4,6}$ components due to the CF potential. However, the extent of that mixing should be no more than 10\% \cite{chen2005}, which makes it to small to induce the strongest $^5$D$_0 \leftrightarrow {}^7$F$_0$ lines.

Moreover, the observation of the $^5$D$_0 \leftrightarrow {}^7$F$_0$ transition is an indication that the Eu$^{3+}$ ion occupies a site with intermediate low symmetry, like $C_{nv}$, $C_n$ or $C_s$ \cite{binnemans1996, binnemans1997}.
Although that transition is often very weak, it is unusually intense in the $\beta$-diketonate, with the Eu$^{3+}$ ion at a site with $C_3$ symmetry \cite{kirby1983}. Unusually high intensities for the $^5$D$_0 \leftrightarrow {}^7$F$_0$ transition are also observed for Eu$^{3+}$ in fluorapatite, hydroxyapatite, oxysulfates, $\alpha$-cordierite, mullite, etc.

Chen \textit{et al.} listed some anomalous Eu$^{3+}$ containing systems, in which very strong ratios of $\frac{I_{00}}{I_{01}}$ are found, where $I_{00}$ is the intensity of $^5$D$_0 \leftrightarrow {}^7$F$_0$ and $I_{01}$ is the intensity of $^5$D$_0 \leftrightarrow {}^7$F$_1$ \cite{chen2005}. Several interesting features can be noted from their list: (i) anomalous CF spectra are often found in those systems in which there are oxygen-compensating sites; and (ii) all the systems with a ratio larger than 20 have $C_s$ symmetry. The most probable explanation for this is that Eu ions, which occupy the $C_s$ position, are surrounded by oxygen atoms from other host groups, and the CF is deformed by O \cite{karbowiak2000}.
This could mean that the presence of oxygen atoms in the host material tends to induce a rather strong $^5$D$_0 \leftrightarrow {}^7$F$_0$ transition. This is the case in the crystals studied in the present article: for example, the composition of lithium borate of Ref.~ \cite{babu2000} is L6BE = 39.5Li$_2$CO$_3$ + 59.5 H$_3$BO$_3$ + 1Eu$_2$O$_3$.

In our models of article I and of the present article, the $^7$F$_0 \leftrightarrow {}^5$D$_0$ transition is allowed. Its line strength is proportional to $X_1$, and its transition amplitude to the CF parameters $A_{1q}$, which tend to increase for lower symmetries. Therefore, it can predict a rather intense transition. With Babu's data \cite{babu2000} in Paper I, the ratio between the theoretical and experimental OSs is equal to 20 in the third-order correction and 7.8 in the second-order one. With the present model, it goes down to 4.4 with or without the host dispersion (the theoretical OS is respectively $6.995 \times 10^{-8}$ and $7.00 \times 10^{-8}$). This improved prediction is certainly due to the inclusion of all eigenvector components in both levels, especially the $^3$P6 one, as mentioned in article I. Still, it is important to mention that, with the data set of Ref.~\cite{kedziorski2007}, the ratio is very large, namely equal to 20.9 (the calculated OS is $3.13 \times 10^{-8}$).

\section{Results for neodymium}
\label{sec:Nd3+}

\subsection{Free-ion calculations}

We have carried out similar calculations for Nd$^{3+}$. Because our model relies on free-ion properties, we start with studying the free-ion energies of the two electronic configurations of Nd$^{3+}$: $4f^3$(odd parity) and $4f^2 5d$ (even parity). Those calculations are described in subsection \ref{subsec:eu-free}: 41 odd-parity and 111 even-parity experimental levels from Ref.~\cite{wyart2007} are included in our fit. Note that the {}``o'' superscript used to designate odd-parity terms is omitted here.

\begin{table*}
\caption{\label{tab:Nd_ground} Comparison between the experimental \cite{wyart2007} and computed values for the levels of 4f$^3$ configuration of Nd$^{3+}$, with total angular momenta from $J=0.5$ to 7.5 and energies up to 30000~cm$^{-1}$, as well as five leading eigenvector components with non-zero percentages. All energy values are in cm$^{-1}$.}
\begin{tabular}{rrrlrlrlrlrlrl}
\\
\toprule
 Exp. & This work & $J$  &  \multicolumn{10}{c} {Eigenvectors with non-zero percentages } & Label\\ \hline
 & & & & & & & & & & & & \\
 0 & 74 & 4.5 & $^4$I & 97.1 \% & $^2$H2 & 2.6 \% & $^2$H1 & 0.3 \% & & & & & $^4$I$_{9/2}$ \\
1897 & 1961 & 5.5 & $^4$I & 99.0 \% & $^2$H2 & 0.9 \% & $^2$H1 & 0.1 \% & & & & & $^4$I$_{11/2}$ \\
3907 & 3975 & 6.5 & $^4$I & 99.6 \% & $^2$K & 0.4 \% & & & & & & & $^4$I$_{13/2}$ \\
5989 & 6075 & 7.5 & $^4$I & 98.8 \% & $^2$K & 1.2 \% & & & & & & & $^4$I$_{15/2}$ \\
11698 & 11746 & 1.5 & $^4$F & 94.3 \% & $^2$D1 & 4.8 \% & $^2$P & 0.3 \% & $^2$D2 & 0.3 \% & $^4$S & 0.3 \% & $^4$F$_{3/2}$ \\
12748 & 12800 & 2.5 & $^4$F & 97.7 \% & $^2$D1 & 2.1 \% & $^2$F2 & 0.1 \% & $^2$F1 & 0.1 \% & & & $^4$F$_{5/2}$ \\
12800 & 13002 & 4.5 & $^2$H2 & 55.7 \% & $^4$F & 13.4 \% & $^2$G1 & 10.9 \% & $^2$H1 & 7.9 \% & $^2$G2 & 7.7 \% & $^2$H$_{9/2}$  \\
13720 & 13692 & 1.5 & $^4$S & 94.5 \% & $^2$P & 4.8 \% & $^4$F & 0.5 \% & $^2$D1 & 0.2 \% & & & $^4$S$_{3/2}$  \\
13792 & 13805 & 3.5 & $^4$F & 93.6 \% & $^2$G1 & 3.7 \% & $^2$G2 & 2.4 \% & $^2$F2 & 0.1\% & $^4$G & 0.1 \% & $^4$F$_{7/2}$ \\
14995 & 15100 & 4.5 & $^4$F & 75.8 \% & $^2$H2 & 19 \% & $^2$H1 & 2.2 \%  & $^2$G1 & 1.6 \% & $^2$G2 & 0.7 \% & $^4$F$_{9/2}$ \\
16162 & 16329 & 5.5 & $^2$H2 & 80.5 \% & $^2$H1 & 12.5 \% & $^4$G & 5.8 \% & $^4$I & 0.9 \% & $^2$I & 0.3 \% & $^2$H$_{11/2}$ \\
17707 & 17544 & 2.5 & $^4$G & 98.6 \% & $^2$F1 & 0.7 \% & $^2$F2 & 0.6 \% & $^4$F & 0.1 \% & & & $^4$G$_{5/2}$ \\
17655 & 17711 & 3.5 & $^4$G & 41.9 \% & $^2$G1 & 30.7 \% & $^2$G2 & 23.1 \% & $^4$F & 4.3 \% & & & $^2$G$_{7/2}$ \\
19541 & 19498 & 3.5 & $^4$G & 57.4 \% & $^2$G1 & 24.3 \% & $^2$G2 & 15.7 \% & $^4$F & 2.0 \% & $^2$F2 & 0.3 \% & $^4$G$_{7/2}$ \\
19970 & 19928 & 4.5 & $^4$G & 75.8 \% & $^2$G1 & 7.2 \% & $^2$G2 & 6.5 \% & $^2$H2 & 6.0 \% & $^4$F & 2.9 \% & $^4$G$_{9/2}$ \\
20005 & 19974 & 6.5 & $^2$K & 98.7 \% & $^2$I & 0.9 \% & $^4$I & 0.4 \% & & & & & $^2$K$_{13/2}$ \\
21493 & 21574 & 4.5 & $^2$G1 & 39.1 \% & $^2$G2 & 26.0 \% & $^4$G & 21.6 \% & $^4$F & 7.8 \% & $^2$H2 & 5.4 \% & $^2$G$_{9/2}$ \\
21701 & 21667 & 1.5 & $^2$D1 & 45.8 \% & $^2$P & 43.6 \% & $^4$S & 3.7 \% & $^4$F & 3.6 \% & $^4$D & 1.6 \% & $^2$D$_{3/2}$ \\
22044 & 22006 & 7.5 & $^2$K & 97.7 \% & $^2$L & 4.1 \% & $^4$I & 1.2 \% & & & & & $^2$K$_{15/2}$ \\
22047 & 21986 & 5.5 & $^4$G & 92.7 \% & $^2$H1 & 4.1 \% & $^2$H2 & 3.1 \% & & & & & $^4$G$_{11/2}$ \\
23789 & 23571 & 0.5 & $^2$P & 94.1 \% & $^4$D & 5.9 \% & & & & & & & $^2$P$_{1/2}$ \\
24333 & 24348 & 2.5 & $^2$D1 & 97.5 \% & $^4$F & 2.1 \% & $^2$D2 & 0.3 \% & $^2$F1 & 0.1 \% & & & $^2$D$_{5/2}$ \\
26761 & 26696 & 1.5 & $^2$P & 48.9 \% & $^2$D1 & 44.5 \% & $^2$D2 & 2.8 \% & $^4$F & 1.5 \% & $^4$S & 1.5 \% & $^2$P$_{3/2}$ \\
29010 & 28958 & 1.5 & $^4$D & 82.0 \% & $^2$D2 & 15.0 \% & $^2$P & 1.6 \% & $^2$D1 & 1.3 \% & & & $^4$D$_{3/2}$ \\
29191 & 29121 & 2.5 & $^4$D & 79.8 \% & $^2$D2 & 17.9 \% & $^2$F2 & 1.1 \% & $^2$F1 & 1.1 \% & $^4$G & 0.1 \% & $^4$D$_{5/2}$ \\
29540 & 29533 & 0.5 & $^4$D & 94.1 \% & $^2$P & 5.9 \% & & & & & & & $^4$D$_{1/2}$ \\

\bottomrule
\end{tabular}
\end{table*}

\begin{figure}
\includegraphics[scale=0.63]{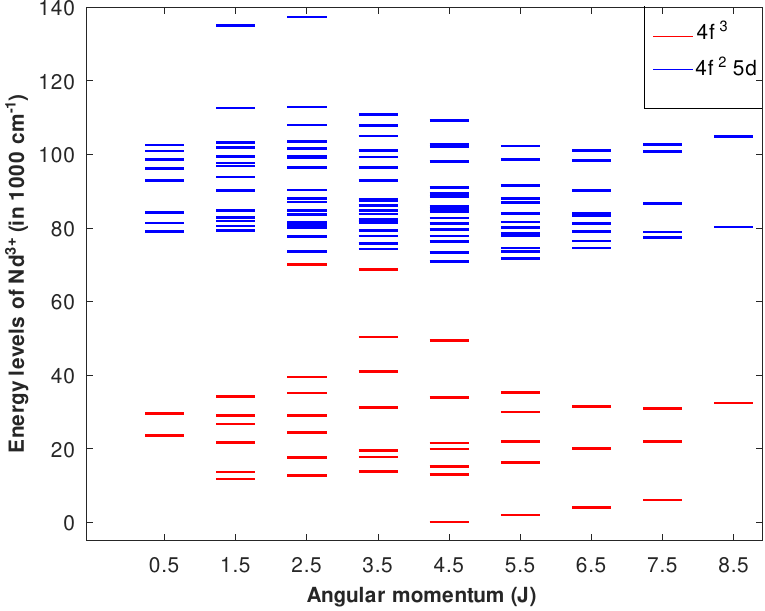}
\caption{\label{fig:Nd_energy_scheme} Energy levels of the 4f$^{3}$ (red) and 4f$^{2} 5d$ (blue) configurations of Nd$^{3+}$ as functions of the electronic angular momentum $J$.
}
\end{figure}

Figure \ref{fig:Nd_energy_scheme} shows the levels computed for both configurations, and their comparison with the data reported in \cite{wyart2007} is shown in table \ref{tab:Nd_ground} for levels below 30000 cm$^{-1}$. We provide also information about our computed eigenvectors, with at most five non-zero percentages. Most of levels are well described by the $LS$ coupling, with leading components above 70~\%. This is less the case for the intermediate $J$-values of 3.5 and 4.5, for which $^2$H, $^2$G, $^4$G and $^4$F manifolds are mixed by the spin-orbit interaction. Note that for the level at 17655 cm$^{-1}$, the leading component is $^4$G with 41.9~\%; but if one adds the two $^2$G manifolds, it yields 53.8~\%. This can lead to some ambiguity when labeling that level. Spin-orbit mixing is also significant between $^2$P and $^2$D manifolds for $J=1.5$.

\begin{table*}
\centering
\caption{\label{tab:Uk_Nd} Comparison between our reduced matrix elements $[U^{(\lambda)}]^2$ for selected transitions of Nd$^{3+}$ and those of Ref.~\cite{carnall1968electronic}.}
\begin{tabular} {@{}lllllll}
\\
\toprule
%
Transition & \multicolumn{2}{c}{[U$^{(2)}]^2$} & \multicolumn{2}{c}{[U$^{(4)}]^2$} & \multicolumn{2}{c}{[U$^{(6)}]^2$}  \\ \cline{1-7}
 & Our & Rep. & Our & Rep. & Our & Rep. \\
\cline{2-3} \cline{4-5} \cline{6-7} \\
$^4$I$_{9/2}\leftrightarrow {}^4$F$_{1/2}$ & 0 & 0 & 0.2297 & 0.2293 & 0.0553 & 0.0549 \\
$^4$I$_{9/2}\leftrightarrow {}^2$H$_{9/2}$ & 0.0089 & 0.0092 & 0.0079 & 0.0080 & 0.1129 & 0.1154 \\

$^4$I$_{9/2}\leftrightarrow {}^4$F$_{7/2}$ $^\mathrm{a}$ & 0.0009 & 0.0010 & 0.0430 & 0.0422 & 0.4238 & 0.4245 \\
$^4$I$_{9/2}\leftrightarrow {}^4$S$_{3/2}$ $^\mathrm{a}$ & 0 & 0 & 0.0026 & 0.0027 & 0.2349 & 0.2352 \\
$^4$I$_{9/2}\leftrightarrow {}^4$F$_{9/2}$ & 0.0009 & 0.0009 & 0.0092 & 0.0092 & 0.0421 & 0.0417 \\
$^4$I$_{9/2}\leftrightarrow {}^4$G$_{5/2}$ & 0.8979 & 0.8979 & 0.4095 & 0.4093 & 0.0356 & 0.0359 \\
$^4$I$_{9/2}\leftrightarrow {}^4$G$_{9/2}$ & 0.0047 & 0.0046 & 0.0603 & 0.0608 & 0.0407 & 0.0406 \\
$^4$I$_{9/2}\leftrightarrow {}^4$G$_{11/2}$ & 0.00001 & $\sim$ 0 & 0.0051 & 0.0053 & 0.0080 & 0.0080 \\
$^4$I$_{9/2}\leftrightarrow {}^2$P$_{1/2}$ & 0 & 0 & 0.0350 & 0.0367 & 0 & 0 \\
$^4$I$_{9/2}\leftrightarrow {}^4$D$_{1/2}$ & 0 & 0 & 0.2603 & 0.2584 & 0 & 0 \\

\bottomrule
\multicolumn{7}{l}{$^\mathrm{a}$ Probable inversion in Table \rom{5} of Ref.~\cite{carnall1968electronic}}
\end{tabular}
\end{table*}

As for Eu$^{3+}$, to check our free-ion eigenvectors, we compare our $[U^{(\lambda)}]^2$ matrix elements with those of Carnall \cite{carnall1968electronic}. Those matrix elements are computed for transitions present in the next subsections. The results are shown in table \ref{tab:Uk_Nd}, showing a very good agreement except for the transitions $^4$I$_{9/2} \leftrightarrow {}^4$S$_{3/2}$ and  $^4$I$_{9/2} \leftrightarrow {}^4$F$_{7/2}$.
By looking closely, we presume that the lines corresponding to those two upper levels in Table \rom{5} of Ref.~\cite{carnall1968electronic} have been interchanged. They are indeed so close in energy that their order is inverted in certain materials. In other words, their absorption peaks overlap, which makes it difficult to correctly identify them. This, for example, happens In the article by Jyothi \textit{et al.} \cite{jyothi2011composition} dedicated to Nd$^{3+}$-doped tellurite and metaborate glasses, where those two transitions are superposed. In this case, the $[U^{(\lambda)}]^2$ matrix elements can be summed to give a single effective transition.

We have also calculated the radial transition integrals $\langle \mathrm{4f} | r^k | \mathrm{5d} \rangle$ necessary for our model. We obtain 1.28773$\,a_0$, -4.10141$\,a_0^3$ and 30.49720$\,a_0^5$ for $k=1$, $k=3$ and $k=5$, respectively,
while the $k=1$ value calculated by Cowan is 1.2877242$\,a_0$.

\begin{figure}
\includegraphics[scale=0.63]{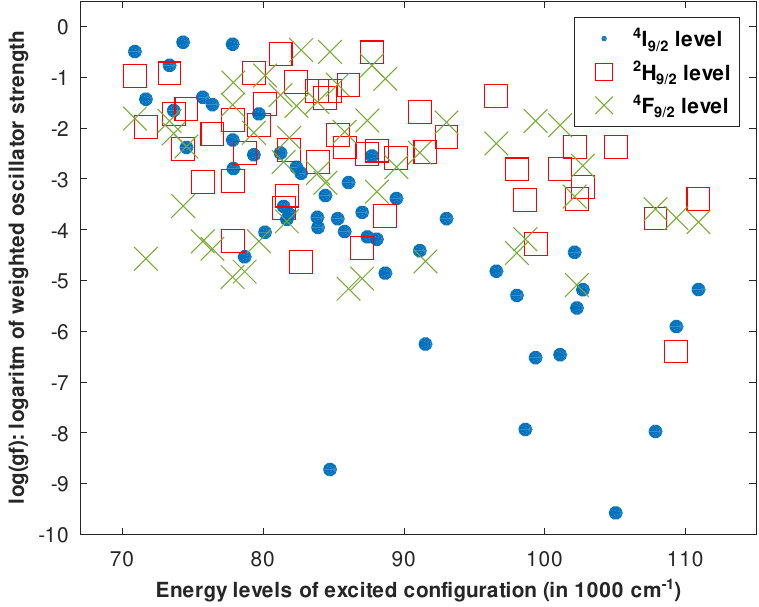}
\caption{\label{fig:Nd_log_gf} Logarithm of the weighted ED oscillator strengths, as functions of the energy of the excited-configuration levels, for transitions implying the $^4$I$_{9/2}$ (blue dots), $^2$H$_{9/2}$ (red squares) and $^4$F$_{9/2}$ (green cross) levels of the ground configuration of Nd$^{3+}$.
}
\end{figure}

Our ability to derive rather simple formulas for the OSs relies in particular on the approximation that all the levels of the first-excited configuration, namely $E_{t,u}$ in Eq.~\eqref{eq:d12-1}, are equal. In order to estimate the best possible value, we search for the range in which the ED coupling involving various levels of the ground configuration is strong. In figure \ref{fig:Nd_log_gf}, we plot the weighted free-ion absorption OSs in log scale, that is the OS multiplied by the degeneracy factor $2J_1+1$ of the lower level. That quantity is indeed proportional to the ED line strength and so to $(\langle \mathrm{4f} | r | \mathrm{5d} \rangle )^2$. For $^4$I$_{9/2}$, the OS shows strong values between 70000 and 80000 cm$^{-1}$, and then it strongly drops. For the two other levels, no such trend is visible. But because the measured transitions in solids  most often involve $^4$I$_J$ levels, we select 75000 cm$^{-1}$ for the energy of excited configuration levels.

\subsection{Nd$^{3+}$ in SrGdGa$_3$O$_7$}

The first set of experimental oscillator strengths is taken from Zhang \textit{et al.} \cite{zhang2010synthesis}, where the authors describe the growth of Nd:SrGdGa$_3$O$_7$ (Nd:SGGM) laser crystal by Czochralski method [Ref \cite{czochralski_usami2011types}] and thermal properties, absorption and emission spectra were measured. In that work, the authors also measure the host refractive index at different wavelengths and fit it using Sellmeier's equation \eqref{eq:Sell_2} with $m=p=1$. Nine absorption transitions were measured in $\sigma$ and $\pi$ polarizations, and the OSs were averaged with factors 2/3 and 1/3 to obtain unpolarized spectra. In Tables \rom{4} and  \rom{5} of Ref.~\cite{zhang2010synthesis}, we take as upper levels those written in the table rows where the OSs are written. In other words, we assume no overlapping transitions. 
Note that, although, our theoretical value of standard JO parameter $\Omega_6$ is different from the one reported by Zhang \textit{et al.}, the general tendency of $\Omega_4 < \Omega_6$ reported in many other articles \cite{choi2005judd, florez2001optical, ma2017spectroscopic, gonccalves2018thermo, rajagukguk2019structural, chanthima2017luminescence}, is conserved.

\begin{table*}
\centering
\caption{\label{tab:JO_Nd} Values of Judd-Ofelt parameters (in 10$^{-20}$ cm$^2$) and $X_k$ (in a.u.), calculated by us (Our), compared with values reported by Zhang \textit{et al.} \cite{zhang2010synthesis} and Chanthima \textit{et al.} \cite{chanthima2017luminescence} (Rep.) for Nd$^{3+}$.}
\begin{tabular}{lccccccccc}
\toprule
%
 & $X_1$ & \multicolumn{2}{c}{$\Omega_2$} & $X_3$ & \multicolumn{2}{c}{$\Omega_4$} & $X_5$ & \multicolumn{2}{c}{$\Omega_6$} \\
 & ($10^{-6}$\,a.u.) & \multicolumn{2}{c}{($10^{-20}$\,cm$^2$)} & ($10^{-6}$\,a.u.) & \multicolumn{2}{c}{($10^{-20}$\,cm$^2$)} & ($10^{-8}$\,a.u.) & \multicolumn{2}{c}{($10^{-20}$\,cm$^2$)} \\ \cline{2-10}
 & & Our & Rep. & & Our & Rep. & & Our & Rep. \\
\cline{3-4} \cline{6-7} \cline{9-10} \\
Nd$^{3+}$:SrGdGa$_3$O$_7$ \cite{zhang2010synthesis} & 2.069 & 1.304 & 1.883 & 1.972 & 5.265 & 4.441 & 3.784 & 7.586 & 2.956 \\
 Nd$^{3+}$:CaO-BaO-P$_2$O$_5$ \cite{chanthima2017luminescence} & 5.035 & 1.547 & 1.09 & 1.859 & 2.850 & 1.97 & 1.702 & 2.388 & 3.37 \\
\bottomrule
\end{tabular}
\end{table*}

With those 9 transitions, the relative standard deviation is 23.78~\% for the present model and 26.61~\% for the standard JO one. Our model is slightly better, but the relative standard deviation remains large. This is certainly because there are several overlapping transitions that our code do not account for. In Ref~\cite{zhang2010synthesis}, the authors obtain a relative standard deviation of 5.4~\%. Those discrepancies are also visible on the JO parameters, as shows Table \ref{tab:JO_Nd}.

Detailed comparisons between experimental and calculated OSs are presented in the upper panel of figure \ref{fig:Nd_jo_histogram} and the left column of Table \ref{tab:Nd_judd_ofelt}. The figure gives a visual insight with histograms of the experimental OSs, and those resulting from our standard JO model and our present extension. The performances of the two models are similar. Table \ref{tab:Nd_judd_ofelt} shows the ratios between experimental OSs and calculated ones with the present model. The agreement is very good for the intense $^4$I$_{9/2} \leftrightarrow {}^4$G$_{5/2}$ transition, which according to Ref.~\cite{zhang2010synthesis} is isolated. On the contrary, the transition $^4$I$_{9/2} \leftrightarrow {}^4$D$_{1/2}$ has a significantly larger experimental OS, certainly due to superimposition with transition peaks with upper states like $^4$D$_{3/2}$, $^4$D$_{5/2}$ and $^2$I$_{11/2}$ as described in the articles of Florez \textit{et al.} \cite{florez2001optical}, Singh \textit{et al.} \cite{singh2014spectroscopic}, in Ma \textit{et al.} \cite{ma2017spectroscopic}, or in Sardar \textit{et al.} \cite{sardar2003judd}. We see the same phenomenon with the transition $^4$I$_{9/2} \leftrightarrow {}^2$H$_{9/2}$: in many articles \cite{sardar2003judd, karunakaran2010structural, florez2001optical, singh2014spectroscopic, wang2011optical}, this transition is reported to be superimposed with a transition with upper state of $^4$F$_{5/2}$.

\subsection{Nd$^{3+}$ in CaO-BaO-P$_2$O$_5$}

\begin{figure}
\includegraphics[scale=0.63]{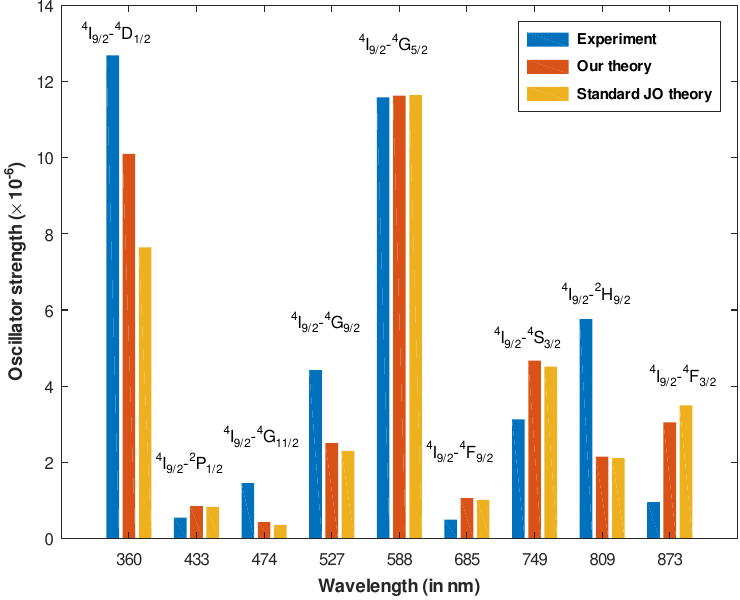}
\includegraphics[scale=0.63]{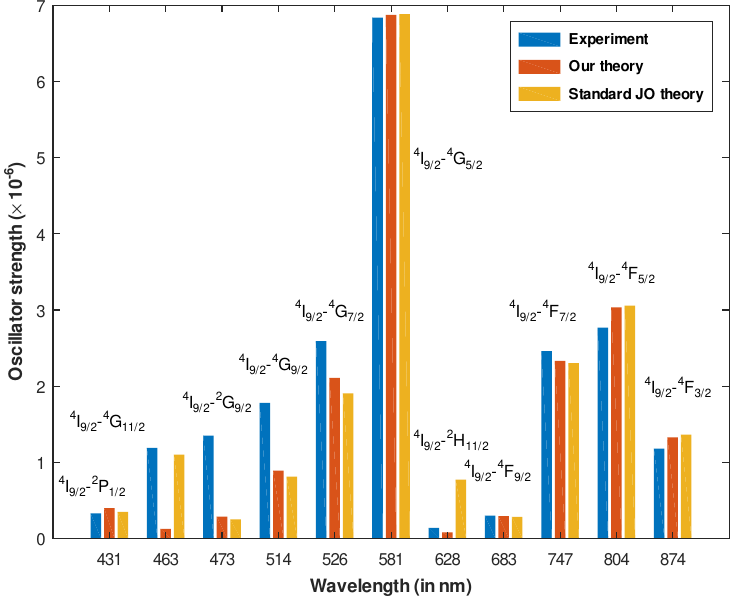}
\caption{\label{fig:Nd_jo_histogram} Comparison between experimental (top panel: \cite{zhang2010synthesis}, bottom panel: \cite{chanthima2017luminescence}) and theoretical oscillator strengths of absorption, plotted as function of the transition wavelength (not at scale). The transitions are labeled with the LS-term quantum numbers of the Nd$^{3+}$ free ion.
}
\end{figure}

We did similar calculations with another set of absorption transitions, reported in Chanthima \textit{et al.}, where the authors do luminescence study and Judd-Ofelt analysis of CaO-BaO-P$_2$O$_5$ glasses doped with Nd$^{3+}$ ions. For this glass we had difficulties to find the Sellmeier parameters, consequently the refractive index is assumed to be constant and equal to 1.556. When 11 transitions are included in the calculations, the relative standard deviation for standard JO calculation is 8.86~\%. The resulting JO parameters are shown in table \ref{tab:JO_Nd}, with a comparison with values reported in the article. The relative standard deviation with the present model is 8.16~\%, a little better than the JO one, and much better than the one obtained with the data of Zhang \textit{et al.}.

\begin{table}
\caption{\label{tab:Nd_judd_ofelt}{Transition labels and ratios between theoretical and experimental oscillator strength for Nd$^{3+}$, when the experimental data for the calculation is taken from \cite{zhang2010synthesis} (left part) and \cite{chanthima2017luminescence} (right part). The last line presents the relative standard deviations for each calculation.}}

\begin{tabular}{ccc}
\toprule
Transition & Nd$^{3+}$:SrGdGa$_3$O$_7$  & Nd$^{3+}$:CaO-BaO-P$_2$O$_5$  \\
Label & Zhang \cite{zhang2010synthesis} & Chanthima \cite{chanthima2017luminescence} \\ \hline
 & & \\
$^4$I$_{9/2} \leftrightarrow {}^4$F$_{3/2}$ & 3.18 & 1.13 \\
$^4$I$_{9/2} \leftrightarrow {}^2$H$_{9/2}$ & 0.37 & \\
$^4$I$_{9/2} \leftrightarrow {}^4$F$_{5/2}$ & & 1.10 \\
$^4$I$_{9/2} \leftrightarrow {}^4$S$_{3/2}$ & 1.49 & \\
$^4$I$_{9/2} \leftrightarrow {}^4$F$_{7/2}$ & & 0.95 \\
$^4$I$_{9/2} \leftrightarrow {}^4$F$_{9/2}$ & 2.13 & 0.98 \\
$^4$I$_{9/2} \leftrightarrow {}^2$H$_{11/2}$ & & 0.59\\
$^4$I$_{9/2} \leftrightarrow {}^4$G$_{5/2}$ & 1.00 & 1.00 \\
$^4$I$_{9/2} \leftrightarrow {}^4$G$_{7/2}$ & & 0.81 \\
$^4$I$_{9/2} \leftrightarrow {}^4$G$_{9/2}$ & 0.57 & 0.50 \\
$^4$I$_{9/2} \leftrightarrow {}^2$G$_{9/2}$ & & 0.21 \\
$^4$I$_{9/2} \leftrightarrow {}^4$G$_{11/2}$ & 0.30 & 0.11 \\
$^4$I$_{9/2} \leftrightarrow {}^2$P$_{1/2}$ & 1.55 & 1.21 \\
$^4$I$_{9/2} \leftrightarrow {}^4$D$_{1/2}$ & 0.80 & \\
\bottomrule
$\sigma/\mathcal{S}_\mathrm{max}$ & 23.78~\% & 8.16~\%
\end{tabular}
\end{table}

The results of calculations are summarized in the bottom panel of figure \ref{fig:Nd_jo_histogram} and in the right part of Table \ref{tab:JO_Nd}. They confirm that the overall agreement is better than for the data set of Zhang and coworkers \cite{zhang2010synthesis}, probably because there are less overlapping transitions. Still, the OSs of the $^4$I$_{9/2} \leftrightarrow {}^2$G$_{9/2}$ and $^4$I$_{9/2} \leftrightarrow {}^4$G$_{11/2}$ transitions are strongly underestimated by our model (as in Table 2 of Ref.~\cite{chanthima2017luminescence}), which may be due to the overlap with the upper level transitions at 22044 and 20005 cm$^{-1}$, respectively.

\section{Results for erbium}
\label{sec:Er3+}

\subsection{Free-ion calculation}

Now we test our model with another ion: Er$^{3+}$. The free-ion calculations have been done with 38 experimental levels of configuration $4f^{11}$ and 58 of $4f^{10} 5d$, which are taken from Meftah \textit{et al.} \cite{meftah2016}. Note that the {}``o'' superscript used to designate odd-parity terms is omitted here.

\begin{figure}
\includegraphics[scale=0.63]{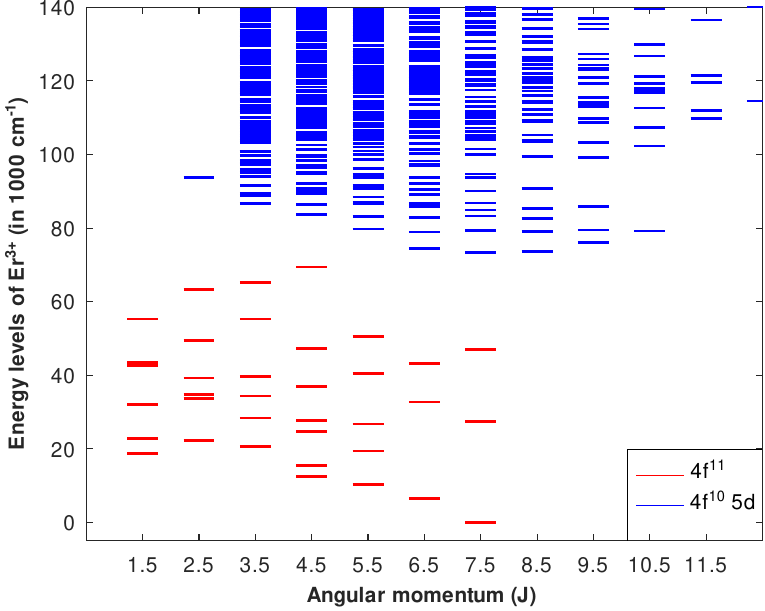}
\caption{\label{fig:Er_energy_scheme} Energy levels of the 4f$^{11}$ (red) and 4f$^{10}$5d (blue) configurations of Er$^{3+}$ as functions of the electronic angular momentum $J$.
}
\end{figure}

\begin{table*}
\caption{\label{tab:Er_ground} Comparison between the experimental \cite{meftah2016} and computed values for the levels of 4f$^{11}$ configuration of Er$^{3+}$, with total angular momenta from $J=1.5$ to 7.5 and energies up to 30000~cm$^{-1}$, as well as five leading eigenvector components with non-zero percentages. The last column gives the level assignment that we retain. All energy values are in cm$^{-1}$.}
\begin{tabular}{rrrlrlrlrlrlrl}
\\
\toprule
 Exp. & This work & $J$  &  \multicolumn{10}{c} {Eigenvectors with non-zero percentages } & Label \\ \hline
& & & & & & & & & & & & \\
0 & -1 & 7.5 & $^4$I & 97.0 \% & $^2$K & 3.0 \% & & & & & & & $^4$I$_{15/2}$ \\
6508 & 6531 & 6.5 & $^4$I & 99.1 \% & $^2$K & 0.8 \% & $^2$I & 0.1 \% & & & & & $^4$I$_{13/2}$ \\
10172 & 10167 & 5.5 & $^4$I & 82.4 \% & $^2$H2 & 14.8 \% & $^4$G & 1.3 \% & $^2$H1 & 1.1 \% & $^2$I & 0.4 \% & $^4$I$_{11/2}$ \\
12469 & 12429 & 4.5 & $^4$I & 53.8 \% & $^2$H2 & 17.6 \% & $^4$F & 12.3 \% & $^2$G1 & 7.7 \% & $^2$G2 & 4.8 \% & $^4$I$_{9/2}$ \\
15405 & 15413 & 4.5 & $^4$F & 59.6 \% & $^4$I & 25.3 \% & $^2$G1 & 8.7 \% & $^2$G2 & 4.8 \% & $^4$G & 0.8 \% & $^4$F$_{9/2}$ \\
- & 18755 & 1.5 & $^4$S & 67.8 \% & $^2$P & 18.6 \% & $^2$D1 & 7.9 \% & $^4$F & 5.5 \% & $^4$D & 0.2 \% & $^4$S$_{3/2}$ \\
19332 & 19343 & 5.5 & $^2$H2 & 48.3 \% & $^4$G & 34.2 \% & $^4$I & 15.0 \% & $^2$H1 & 2.1 \% & $^2$I & 0.3 \% & $^2$H$_{11/2}$ \\
- & 20690 & 3.5 & $^4$F & 92.3 \% & $^2$G1 & 4.6 \% & $^2$G2 & 2.5 \% & $^2$F2 & 0.3 \% & $^2$F1 & 0.2 \% & $^4$F$_{7/2}$ \\
- & 22294 & 2.5 & $^4$F & 83.9 \% & $^2$D1 & 13.0 \% & $^2$D2 & 2.0 \% & $^2$F2 & 0.5 \% & $^4$D & 0.2 \% & $^4$F$_{5/2}$ \\
- & 22708 & 1.5 & $^4$F & 62.6 \% & $^2$D1 & 20.1 \% & $^4$S & 16.9 \% & $^2$P & 0.4 \% & & & $^4$F$_{3/2}$ \\
24736 & 24736 & 4.5 & $^4$F & 24.3 \% & $^2$G1 & 19.0 \% & $^2$H2 & 16.6 \% & $^2$G2 & 14.9 \% & $^4$I & 12.4 \% & $^2$G$_{9/2}$ \\
26708 & 26739 & 5.5 & $^4$G & 61.6 \% & $^2$H2 & 25.5 \% & $^2$H1 & 1.5 \% & $^4$I & 2.4 \% & & & $^4$G$_{11/2}$ \\
27767 & 27738 & 4.5 & $^4$G & 79.5 \% & $^2$H2 & 14.5 \% & $^4$I & 4.7 \% & $^2$H1 & 0.8 \% & $^2$G2 & 0.4 \% & $^4$G$_{9/2}$ \\
- & 27353 & 7.5 & $^2$K & 90.9 \% & $^2$L & 60.1 \% & $^4$I & 3.0 \% & & & & & $^2$K$_{15/2}$ \\
- & 28311 & 3.5 & $^4$G & 41.6 \% & $^2$G1 & 26.6 \% & $^2$G2 & 23.3 \% & $^4$F & 3.9 \% & $^2$F2 & 2.2 \% & $^2$G$_{7/2}$ \\
\bottomrule
\end{tabular}
\end{table*}

Figure \ref{fig:Er_energy_scheme} shows the levels computed for the 4f$^{11}$ and 4f$^{10}$5d configurations up to 120000 cm$^{-1}$. It shows in particular  a large density in the excited configuration, which is due to the four vacancies in the 4f shell.
For levels up to 30000 cm$^{-1}$, Table \ref{tab:Er_ground} shows a comparison between experimental and theoretical energies, which happens to be very good. Compared to neodymium, the density of ground-configuration levels is smaller for erbium, which reduces the probability of overlapping transitions.
Table \ref{tab:Er_ground} also presents up to five eigenvector components with non-zero percentages. The $LS$ coupling scheme applies to a lesser extent than for neodymium, which is due to the larger spin-orbit interaction. For the levels with calculated energies of 24736 and 28311 cm$^{-1}$, labeling is not trivial. For the former, the sum of $^2$G components gives the largest contribution of 33.9~\%, and so we retain the label $^2$G$_{9/2}$. For the latter, the sum of $^2$G components, equal to 49.9~\% exceeds the $^4$G one: therefore we retain the label $^2$G$_{7/2}$ (see last column of Table \ref{tab:Er_ground}).

\begin{table*}
\centering
\caption{\label{tab:Uk_Er} Comparison between our reduced matrix elements $[U^{(\lambda)}]^2$ for selected transitions of Er$^{3+}$ and those of Ref.~\cite{carnall1968electronic}.}
\begin{tabular} {@{}lllllll}
\\
\toprule
%
Transition & \multicolumn{2}{c}{$[U^{(2)}]^2$} & \multicolumn{2}{c}{$[U^{(4)}]^2$} & \multicolumn{2}{c}{$[U^{(6)}]^2$}  \\ \cline{1-7}
 & Our & Rep. & Our & Rep. & Our & Rep. \\
\cline{2-3} \cline{4-5} \cline{6-7} \\
$^4$I$_{15/2}\leftrightarrow {}^4$I$_{13/2}$ & 0.0195 & 0.0195 & 0.1173 & 0.1173 & 1.4304 & 1.4316 \\
$^4$I$_{15/2}\leftrightarrow {}^4$I$_{11/2}$ & 0.0275 & 0.0282 & 0.0002 & 0.0003 & 0.3983 & 0.3953 \\
$^4$I$_{15/2}\leftrightarrow {}^4$I$_{9/2}$ & 0 & 0 & 0.1504 & 0.1733 & 0.0053 & 0.0099 \\
$^4$I$_{15/2}\leftrightarrow {}^4$F$_{9/2}$ & 0 & 0 & 0.5581 & 0.5581 & 0.4643 & 0.4643 \\
$^4$I$_{15/2}\leftrightarrow {}^4$S$_{3/2}$ & 0 & 0 & 0 & 0 & 0.2191 & 0.2191 \\
$^4$I$_{15/2}\leftrightarrow {}^2$H$_{11/2}$ & 0.6922 & 0.7125 & 0.3973 & 0.4125 & 0.0865 & 0.0925 \\
$^4$I$_{15/2}\leftrightarrow {}^4$F$_{7/2}$ & 0 & 0 & 0.1467 & 0.1469 & 0.6272 & 0.6266 \\
$^4$I$_{15/2}\leftrightarrow {}^4$F$_{5/2}$ & 0 & 0 & 0 & 0 & 0.2222 & 0.2232 \\
$^4$I$_{15/2}\leftrightarrow {}^2$G$_{9/2}$ & 0 & 0 & 0.0217 & 0.0189 & 0.2215 & 0.2256\\
$^4$I$_{15/2}\leftrightarrow {}^4$G$_{11/2}$ & 0.9391 & 0.9183 & 0.5381 & 0.5262 & 0.1215 & 0.1235 \\
$^4$I$_{15/2}\leftrightarrow {}^2$G$_{7/2}$ & 0 & 0 & 0.0175 & 0.0174 & 0.1158 & 0.1163 \\
$^4$I$_{15/2}\leftrightarrow {}^4$G$_{9/2}$ & 0 & 0 & 0.2380 & 0.2416 & 0.1293 & 0.1235 \\

\bottomrule
\end{tabular}
\end{table*}

Table \ref{tab:Uk_Er} shows results for $[U^{(\lambda)}]^2$ matrix elements calculated with our eigenvectors, in comparison with values reported in the article of Carnall \cite{carnall1968electronic}. It shows an overall good agreement, except for $[\langle {}^4\mathrm{I}_{15/2} \|U^{(6)}\| {}^4\mathrm{I}_{9/2} \rangle]^2$ that we find almost twice as small as Carnall.

\begin{figure}
\includegraphics[scale=0.63]{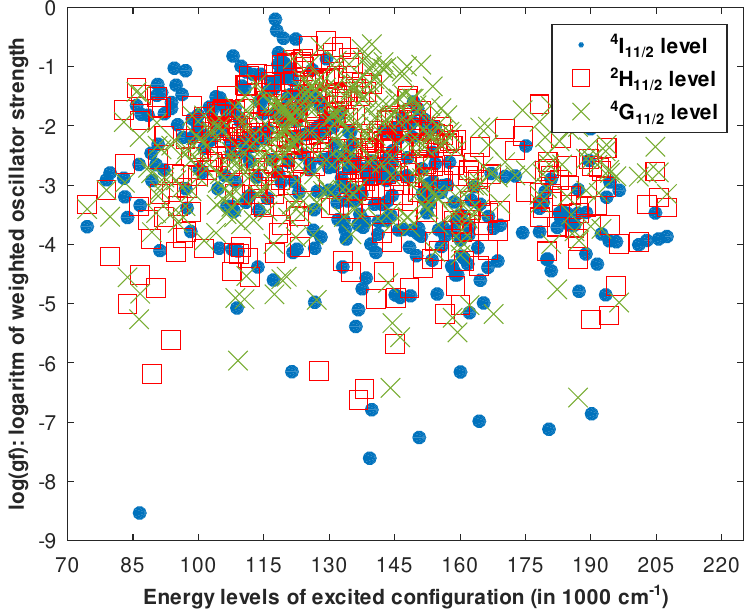}
\caption{\label{fig:Er_log_gf} Logarithm of the free-ion weighted ED oscillator strengths, as functions of the energy of the excited-configuration levels, for transitions implying the $^4$I$_{11/2}$ (blue dots), $^2$H$_{11/2}$ (red squares) and $^4$G$_{11/2}$ (green cross) levels of the ground configuration of Er$^{3+}$.
}
\end{figure}

We have also calculated the matrix elements $\langle n' l' | r^k | n l \rangle$ for Er$^{3+}$, where $n l=4 f$ and $n' l'=5 d$. We obtain 0.96441$\,a_0$, -2.37459$\,a_0^3$ and 14.24536$\,a_0^5$ for $k=1$, 3 and 5, respectively, while the value calculated for this matrix element by Cowan codes is 0.9644014. Based on $\langle \mathrm{4f} | r | \mathrm{5d} \rangle$, we plot on Figure \ref{fig:Er_log_gf} the logarithm of the weighted free-ion oscillator strengths as functions of the excited-configuration level energy, for transitions involving three $J=11/2$ levels of the ground configuration.
It shows that the energy band with strong transitions, in other words, the strong-coupling window for Er$^{3+}$ is between 115000 and 160000 cm$^{-1}$. Therefore, as the excited-configuration energy $E_{t,u}$ in Eq.~\eqref{eq:d12-1}, we do not take the center-of-gravity energy of the excited-configuration, but a value of 145000 cm$^{-1}$.

\subsection{Er$^{3+}$ in Lu$_3$Ga$_5$O$_{12}$}

As a first set of OSs, we take the article by Liu \textit{et al.} \cite{liu2021sellmeier}, where the authors report growth, refractive index dispersion, optical absorption and Judd-Ofelt spectroscopic properties of Er$^{3+}$-doped lutetium gallium garnet (Lu$_3$Ga$_5$O$_{12}$) single-crystal.
A fit of their measured refractive index with Eq.~\eqref{eq:Sell_3} gives $n_0=1$, $A=2.72452$ and $B=0.0172907~\mu$m$^2$. Following the discussion of Table \ref{tab:Er_ground}, we cautiously examine the transition labels of the article.

For the transition labeled $^4$I$_{15/2} \leftrightarrow {}^2$H$_{9/2}$, Liu \textit{et al.} report a wavelength of 410\,nm, which corresponds to the energy level close to 245000~cm$^{-1}$. In our free-ion calculations (see Table \ref{tab:Er_ground}) the dominant eigenvector component of this level is 24.3~\% $^4$F, but its largest $LS$ term is $^2$G.

We exclude from the fit the overlapping transitions $^4$I$_{15/2} \leftrightarrow {}^4$F$_{5/2,3/2}$, as well as the transition $^4$I$_{15/2} \leftrightarrow {}^4$I$_{9/2}$ because we obtain a very small ratio of $\sim 10^{-2}$ between calculated and experimental OSs.

\begin{table}
\caption{\label{tab:Er_judd_ofelt}{Transition labels and ratios between theoretical and experimental line strength for Er$^{3+}$, when the experimental data for the calculation is taken from \cite{liu2021sellmeier} and \cite{piao2019optical}. The last line presents the relative standard deviations for each calculation.}}

\begin{tabular}{cccc}
\toprule
Transition & Er$^{3+}$:Lu$_3$Ga$_5$O$_{12}$ & Er$^{3+}$:SrGdGa$_3$O$_7$ \\
Label & Liu \cite{liu2021sellmeier} & Piao \cite{piao2019optical} \\ 
\hline
 & & \\
$^4$I$_{15/2} \leftrightarrow {}^4$I$_{13/2}$ & 0.87 & 0.88 \\
$^4$I$_{15/2} \leftrightarrow {}^4$I$_{11/2}$ & 0.90 & 1.52 \\ 
$^4$I$_{15/2} \leftrightarrow {}^4$I$_{9/2}$ & & 0.93 \\
$^4$I$_{15/2} \leftrightarrow {}^4$F$_{9/2}$ & 0.80 & 0.97 \\
$^4$I$_{15/2} \leftrightarrow {}^4$S$_{3/2}$ & 1.33 & 1.14 \\
$^4$I$_{15/2} \leftrightarrow {}^2$H$_{11/2}$ & 0.85 & 0.84 \\
$^4$I$_{15/2} \leftrightarrow {}^4$F$_{7/2}$ & 2.83 & 1.53 \\
$^4$I$_{15/2} \leftrightarrow {}^4$F$_{5/2}$ & & 1.29 \\
$^4$I$_{15/2} \leftrightarrow {}^2$G$_{9/2}$ & 2.34 & 1.67 \\
$^4$I$_{15/2} \leftrightarrow {}^4$G$_{11/2}$ & 1.10 & 1.09 \\
$^4$I$_{15/2} \leftrightarrow {}^2$G$_{7/2}$ & 3.07 & & \\
$^4$I$_{15/2} \leftrightarrow {}^4$G$_{9/2}$ & & 0.93 \\
\bottomrule
$\sigma/\mathcal{S}_\mathrm{max}$ & 13.36~\% & 7.48~\%
\end{tabular}
\end{table}

\begin{figure}
\includegraphics[scale=0.63]{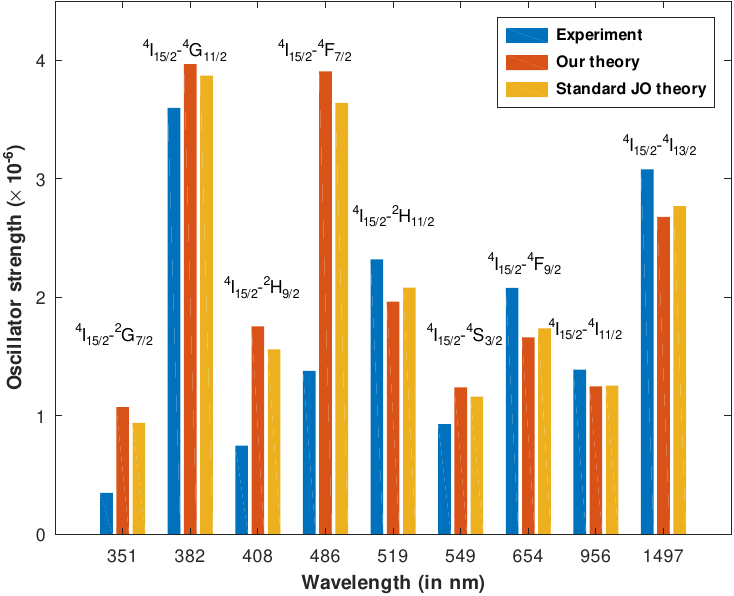}
\includegraphics[scale=0.63]{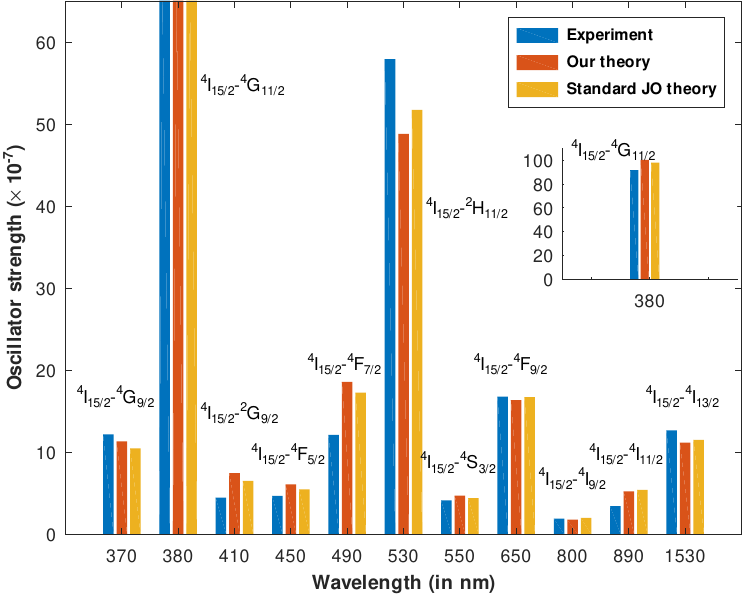}
\caption{\label{fig:Er_jo_histogram} Comparison between experimental (top panel: \cite{liu2021sellmeier}, bottom panel: \cite{piao2019optical}) and theoretical oscillator strengths of absorption, plotted as function of the transition wavelength (not at scale). The transitions are labeled with the LS-term quantum numbers of the Er$^{3+}$ free ion.
}
\end{figure}

The relative standard deviation with the JO model is 11.49~\%; the one with our model is 13.36~\%. The better performance of the standard JO model is visible for each transition of the upper panel of Figure \ref{fig:Er_jo_histogram}. Regarding the fitted parameters, we obtain negative values of $X_3$ and $\Omega_4$, which is abnormal since all parameters should be positive. The $\Omega_4$ value of Liu \textit{et al.} \cite{liu2021sellmeier}, although positive, is small compared to the other $\Omega_\lambda$. Their $\Omega_2$ and $\Omega_6$ strongly differ from ours.

\subsection{Er$^{3+}$ in SrGdGa$_3$O$_7$}

The second set of experimental OSs is taken from the article of Piao \textit{et al.}, where the authors describe optical and Judd-Ofelt spectroscopic study of Er$^{3+}$-doped strontium gadolinium gallium garnet single‐crystal \cite{piao2019optical}.
For this second set of absorption data we did the calculations once, assuming the refractive index is constant  and equal to 1.81014 for all wavelength values, since it was impossible to find values for Sellmeier coefficients for the crystal investigated in the article of Piao \textit{et al.} \cite{piao2019optical}.

The level identification for this data set was a bit delicate. This is especially the case for transitions $^4$I$_{15/2}\leftrightarrow {}^2$H$_{9/2}$ and $^4$I$_{15/2}\leftrightarrow {}^2$G$_{9/2}$ as identified in the article of Piao \textit{et al}. However our free-ion calculations show that the first one should rather be identified as $^4$I$_{15/2}\leftrightarrow {}^2$G$_{9/2}$, and the second as $^4$I$_{15/2}\leftrightarrow {}^4$G$_{9/2}$. This is confirmed by the fact that the peak of the first transition is at 410 nm, which corresponds to the energy level value of ~24300 cm$^{-1}$, having a first term of $^4$F with 24.3\% and two terms of $^2$G with 19.0\% and 14.9\% percentages, making the term $^2$G a dominant one with a percentage of 33.9\%. The identification is possible because this level has $^2$H term with a 16.6\% (see table \ref{tab:Er_ground}).

We have a tricky situation for the second absorption band as well, which in the article of Piao \textit{et al.} is indicated to be at 370 nm, corresponding to the energy level of $\sim$27000 cm$^{-1}$. Our free-ion calculations show that the first and dominant $LS$ term for this level is 79.5\% $^4$G, but it has a $^2$G term with 0.4\%, which makes the identification somehow possible (see table \ref{tab:Er_ground}). It our calculations, however, we will use the labeling corresponding to our free-ion calculation results.

\begin{table*}
\centering
\caption{\label{tab:JO_Er} Values of Judd-Ofelt parameters (in 10$^{-20}$ cm$^2$) for Er$^{3+}$, compared with values reported in Liu \textit{et al.} \cite{liu2021sellmeier} and Piao \textit{et al.} \cite{piao2019optical}.}
\scalebox{0.89} {
\begin{tabular}{lccccccccc}
\toprule
%
 & $X_1$ & \multicolumn{2}{c}{$\Omega_2$} & $X_3$ & \multicolumn{2}{c}{$\Omega_4$} & $X_5$ & \multicolumn{2}{c}{$\Omega_6$} \\
 & ($10^{-5}$\,a.u.) & \multicolumn{2}{c}{($10^{-20}$\,cm$^2$)} & ($10^{-6}$\,a.u.) & \multicolumn{2}{c}{($10^{-20}$\,cm$^2$)} & ($10^{-7}$\,a.u.) & \multicolumn{2}{c}{($10^{-20}$\,cm$^2$)} \\ \cline{2-10}
 & & Our & Rep. & & Our & Rep. & & Our & Rep. \\
\cline{3-4} \cline{6-7} \cline{9-10} \\
Er$^{3+}$:Lu$_3$Ga$_5$O$_{12}$ \cite{liu2021sellmeier} & 17.11 & 2.095 & 0.89 $\pm$ 0.16 & -13.82 & -0.5706 & 0.16 $\pm$ 0.10 & 10.71 & 4.296 & 1.85 $\pm$ 0.25 \\
Er$^{3+}$:SrGdGa$_3$O$_7$ \cite{piao2019optical} & 19.17 & 2.792 & 2.46 & 11.97 & 0.8883 & 1.24 & 2.387 & 0.9541 & 0.51 \\

\bottomrule
\end{tabular}
}
\end{table*}

When 11 transitions were included the standard deviation with the JO model is 5.63~\%, with our model it is 7.48~\%. Table \ref{tab:JO_Er} shows results for JO parameters $\Omega_\lambda$, in comparison with values reported in Piao \textit{et al.} \cite{piao2019optical} as well as the fitting parameters $X_k$, which are all positive, and follow the trend $X_5 < X_3 < X_1$. Table \ref{tab:Er_judd_ofelt} and figure \ref{fig:Er_jo_histogram} show, unlike the previous data set, a good match between the OSs of the $^4$I$_{15/2} \leftrightarrow {}^4$I$_{9/2}$ transition.

\section{Conclusions}
\label{sec:conclusions}

In this article, we propose an extension of the Judd-Ofelt model, to describe the absorption or emission line intensities of solids doped with lanthanide trivalent ions. We give expressions of the transition line strengths in which the properties of the Ln$^{3+}$ impurity are fixed parameters accurately calculated with free-ion spectroscopy, and the crystal-field parameters are adjusted by least-square fitting.
Compared to our previous work \cite{hovhannesyan2022transition}, the spin-orbit interaction within the first excited configuration 4f$^{w-1}$5d is described in a perturbative way, whereas it is exactly taken into account in the ground configuration 4f$^w$. For the free-ion levels of this configuration, all the eigenvector components are presently included in the calculation. The wavelength dependence of the refractive index of the host material is also accounted for by means of the Sellmeier equation. The code implementing our model and examples with the data sets used in this article can be found on GitLab \cite{gitlab-joso}.

We have tested the validity of our model on three ions, Eu$^{3+}$, Nd$^{3+}$ and Er$^{3+}$, each hosted in two materials. We have compared our free-ion energies with those available in the literature, and our matrix elements of the unit-tensor operators $[U^{(\lambda)}]^2$ with the values reported in the articles of Carnall \cite{carnall1989, carnall1968electronic}. Using those matrix elements, we have calculated the $\Omega_\lambda$ parameters of the standard Judd-Ofelt theory and compared them with the values reported in the articles from which we took the experimental oscillator strengths used for our fitting procedure. Finally, we compare the performances of our model with those of the standard Judd-Ofelt one.

Our model shows better results in the case of Eu$^{3+}$: not only it allows for interpreting more transitions that the standard Judd-Ofelt model, but it also reproduces more accurately the other oscillator strengths. For the two other ions, in one data set, we obtain comparable performances. But for one data set of Nd$^{3+}$ \cite{zhang2010synthesis}, we observe large discrepancies that we expect to come from overlapping transitions involving close excited levels. To solve this problem, we will add in our code the possibility to treat such situations. In one data set of Er$^{3+}$ \cite{liu2021sellmeier}, we observe some negative fitting parameters, whereas they are supposed to be positive. That abnormal situation is all the more difficult to interpret that the $\Omega_4$ parameter published in Ref.~\cite{liu2021sellmeier}, though positive, is small compared to other parameters. 

The oscillator strengths measured in Ref.~\cite{zhang2010synthesis} separate $\sigma$ and $\pi$ polarizations, giving rather different values. As a prospect, we plan to treat transitions with polarized light or between individual ion-crystal sublevels. This will be possible in our model, because the only fitted parameters are the crystal-field ones. This can open the possibility to model the spectroscopic properties of Ln$^{3+}$-doped nanometer-scale host materials \cite{chacon2020}.

\section*{Acknowledgements}
\label{sec:acknowledgements}

We acknowledge support from the NeoDip project (ANR-19-CE30-0018-01 from ``Agence Nationale de la Recherche''). M.L. also acknowledges the financial support of {}``R{\'e}gion Bourgogne Franche Comt{\'e}'' under the projet 2018Y.07063 {}``Th{\'e}CUP''.
Calculations have been performed using HPC resources from DNUM CCUB (Centre de Calcul de l'Universit\'e de Bourgogne).

\bibliographystyle{model1-num-names}


\end{document}